# Nanoserpents: Graphene Nanoribbons Motion on Two-Dimensional Hexagonal Materials

Wengen Ouyang, Davide Mandelli, Michael Urbakh,* Oded Hod

*School of Chemistry, and The Sackler Center for Computational Molecular and Materials Science, Tel Aviv University, Tel Aviv 6997801, Israel.*

ABSTRACT

We demonstrate snake-like motion of graphene nanoribbons atop graphene and hexagonal boron nitride (*h*-BN) substrates using fully-atomistic non-equilibrium molecular dynamics simulations. The sliding dynamics of the edge-pulled nanoribbons is found to be determined by the interplay between in-plane ribbon elasticity and interfacial lattice mismatch. This results in an unusual dependence of the friction-force on the ribbon's length, exhibiting an initial linear rise that levels-off above a junction dependent threshold value dictated by the pre-slip stress distribution within the slider. As part of this letter, we present the LAMMPS implementation of the registry-dependent interlayer potentials for graphene, *h*-BN, and their heterojunctions that were used herein, which provide enhanced performance and accuracy.

**Keywords**: nanoribbons, graphene, hexagonal boron nitride (*h*-BN), registry-dependent interlayer potential, stress distribution, nanoscale friction, LAMMPS.

## Introduction

Two-dimensional (2D) layered materials such as graphene, hexagonal boron nitride (*h*-BN), molybdenum disulfide ($MoS_2$), and tungsten disulfide ($WS_2$) have attracted great scientific and technological interest due to their unique electronic,[1-3] mechanical,[4-6] and frictional properties.[7-14] In recent years, much attention has been paid to heterogeneous layered materials junctions that may exhibit diverse physical properties as well as enhanced performance over their homogeneous counterparts.[15-17] For instance, recent studies show that graphene/*h*-BN heterostructures may present desired electronic properties[18, 19] as well as robust superlubricity.[20]

Further control over the physical properties of 2D layered materials can be gained via tuning their lateral dimensions. To this end, the aspect-ratio of graphene nanoribbons (GNRs),[21-24] has been long considered as a handle to control their electronic properties.[25-29] Recently, GNRs have also been shown to exhibit ultra-low friction when deposited on gold surfaces.[30-32] This suggests that, when deposited on 2D hexagonal layered materials, where interfacial incommensurability can be controlled, GNRs' motion should exhibit rich behavior.

In the present letter, we consider the motion of edge-driven graphene nanoribbons atop graphene and *h*-BN substrates. Using fully-atomistic molecular dynamics (MD) simulations, we find that the intricate interplay between in-plane ribbon elasticity and interfacial registry results in unique anisotropic snake-like motion. Furthermore, a non-linear dependence of the friction force on the ribbons' length is predicted, where an initial increase is followed by the saturation of friction above a junction-dependent characteristic length, which is determined by the interfacial pre-slip stress distribution within the slider.

In order to allow for the elaborate MD simulations undertaken herein, we provide an efficient LAMMPS implementation of the anisotropic interlayer potentials (ILP) for graphene, *h*-BN, and their heterostructures (see Sections 1-3 of the Supporting Information (SI)).[33-35] To enhance the reliability of our calculations we further refine the ILP's parameterization thus providing a balanced description of the interlayer interactions at both low and high-pressure regimes.

## Methods

### Model System

Our simulated model system consists of an armchair GNR of fixed width (~0.7 nm) and different lengths in the range of 4-60 nm sliding atop rigid graphene or *h*-BN monolayer substrates (see Figure 1). The GNRs' edges are passivated by hydrogen atoms [30] to avoid peripheral C–C bond reconstruction,[36, 37] that may influence friction. The GNRs are initially placed atop the graphene or $h$-BN substrates in three different orientations aligning their long axis parallel to the (i) armchair and (ii) zigzag directions of the hexagonal surfaces, as well as (iii) 45° in between them.

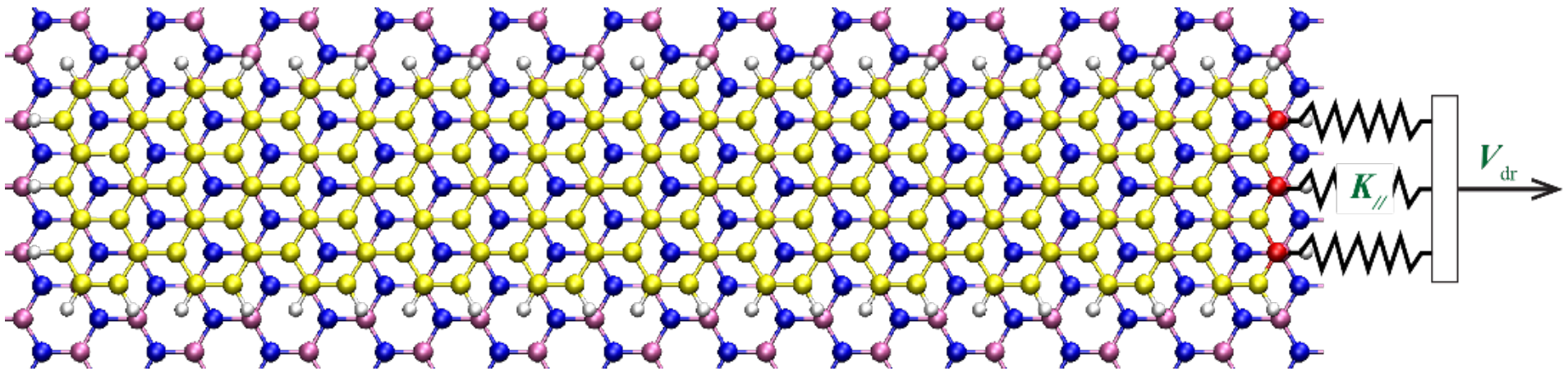

***Figure 1**. A schematic representation of the simulation setup. A graphene nanoribbon deposited over an h-BN substrate monolayer is driven by a stage moving at constant velocity $V_{\mathrm{dr}}$ via springs of stiffness $K_{\|}$ connected to the three rightmost carbon atoms (red spheres). Mauve, blue, yellow, and grey spheres represent boron, nitrogen, carbon, and hydrogen atoms, respectively.*

The intra-layer C-C and C-H interactions within the GNRs were computed via the REBO force field,[38] augmented with a torsion term, which was proposed to improve the description of the mechanical properties of small hydrocarbon molecules.[39] Test simulations performed neglecting the torsion term yielded quantitatively similar results (see Section 4 of the SI).

The interlayer interactions between the GNRs and the two substrates were described via the registry-dependent ILP,[33-35] which we implemented in the LAMMPS[40] suite of codes. We present a refined parametrization of the ILP and the Kolmogorov Crespi (KC) Potential, which provides a balanced treatment of the interlayer interactions in the low and high normal loads regime characterized by interlayer spacing near and below the equilibrium value, respectively. The fitting procedure, the final sets of parameters, the results of several test simulations, and the comparison with the previous parametrizations are discussed in full details in the SI (see Sections 1-3). The results presented herein have been obtained using the ILP parameter set presented in Table S1 of the SI.

<u>Simulation Protocol</u>

All simulations were performed adopting the following protocol. First, we generate the starting configurations of the GNRs' structures via geometry optimization. This is done using the FIRE algorithm,[41] as implemented in LAMMPS,[40] setting a threshold force value of $10^{-6}$ eV/Å. Sliding friction simulations are then carried out by attaching the three rightmost carbon atoms of the GNR (red spheres in Figure 1), via springs of constant $K_{\|}$ in the lateral directions, to a stage of position $\boldsymbol{r}^{stage}(t)$ that is moving along the substrate's armchair axis at constant velocity $V_{\mathrm{dr}}$. The stiffness of the external springs is chosen to be $K_{\parallel} = 3.33\ \mathrm{N/m}$, resulting in an overall effective spring constant of 10 N/m, close to the typical values used in friction force microscopy (FFM) experiments.[42] Due to numerical limitations, the pulling velocity is chosen to be $V_{\mathrm{dr}} = 1\ \mathrm{m/s}$, which is significantly higher than the typical experimental values. Nevertheless, it is sufficiently low to allow for simulating the experimentally observed stick-slip behavior, which is a key feature in the investigated phenomena.

Damped dynamics is applied to avoid junction heating using the following equation of motion:

$$m_i\ddot{\boldsymbol{r}}_i = -\boldsymbol{\nabla}_i\left(V^{\mathrm{inter}} + V^{\mathrm{intra}}\right) - \sum_{\alpha=x,y,z}\eta_\alpha(z_i)m_i\dot{\boldsymbol{r}}_{i,\alpha} + K_{\parallel}(\boldsymbol{r}_i - \boldsymbol{r}^{stage})\delta_{i,i_{\mathrm{edge}}}, \tag{1}$$

where $m_i$ is the mass of atom $i$, $\boldsymbol{r}_i$ is its position, and $V^{\mathrm{inter}}$ and $V^{\mathrm{intra}}$ are the interlayer and intra-layer interaction potentials, respectively. The second term in eq 1 represents viscous damping applied in all directions $\alpha = x, y, z$ to all GNR atoms, while the last term is the driving spring force, which is applied only to the three rightmost edge atoms in the lateral directions (see Figure 1).

The damping coefficients, $\eta_{x,y,z}$, implicitly account for the dissipation of kinetic energy of the GNR into the microscopic degrees of freedom of the substrate. These are dynamically varied according to the following exponential function:[43-45] $\eta_\alpha(z_i) = \eta_\alpha^0\exp\left(1 - z_i/d_{\mathrm{eq}}\right)$, where $z_i$ is the z coordinate of atom $i$ measured with respect to the substrate surface.

In the case of graphene substrate, the value of $d_{\mathrm{eq}}$ is set equal to the density functional theory (DFT) reference equilibrium distance of a graphene bilayer at the energetically optimal AB stacking mode, $d_{\mathrm{eq}} = 3.4$ Å (see Section 2 in the SI).[35] In the case of *h*-BN substrate, $d_{\mathrm{eq}}$ is set equal to the DFT reference equilibrium distance of an artificially commensurate graphene/*h*-BN bilayer at the lattice spacing of 1.43 Å and optimal C-stacking mode, $d_{\mathrm{eq}} = 3.3$ Å (see Section 2 in the SI).[34] The results presented in the main text have been obtained using $\eta_x^0 = \eta_y^0 = \eta_z^0 = 1\ \mathrm{ps}^{-1}$.[46] We checked that the qualitative nature of the simulations results is independent of the choice of $\eta_\alpha^0$ within a broad range of values (see Section 5 of the SI).

A fixed time step of 1 fs was used throughout the simulations. To check for convergence of the results with respect to the time-step we made sensitivity tests by reducing the time step by a factor of 4 leading to practically the same results (see Section 6 of the SI). Since all simulations presented herein were performed at zero temperature, we further evaluated the effects of coupling to a thermal bath via a Langevin thermostat set to room temperature (300 K). The results (presented in Section 7 of the SI) show qualitatively similar frictional behavior.

The time-averaged friction-forces have been calculated as $\langle F_{\mathrm{K}}\rangle = \langle 3K_{||}\left(V_{\mathrm{dr}}t - X_{\mathrm{edge}}\right)\rangle$, where $X_{\mathrm{edge}} = \sum_{i,edge=1}^{3} x_{i,edge}/3$ is the mean position of the nanoribbon's edge atoms along the pulling

direction and $\langle\cdot\rangle$ denotes a steady-state time average. The statistical errors have been estimated using ten different datasets for every system considered, each taken over a time interval of 1 ns.

Stresses are calculated by dividing the global stress tensor, calculated by LAMMPS and given in units of bar $\cdot$ Å$^3$,[47] by the volume associated with a carbon atom. The latter is evaluated as $3\sqrt{3}a_{\mathrm{CC}}^2 \cdot h/4$, where $a_{\mathrm{CC}} = 1.3978$ Å is the equilibrium C-C distance and $h = 3.35$ Å is the effective thickness associated with the GNR, which we fixed to be equal to the equilibrium interlayer distance of graphite.

## Results and Discussion

### Aligned Interfaces

We start by investigating the dependence of the static and kinetic friction forces on the ribbon's length ($L_{\mathrm{GNR}}$) for aligned junctions, where the armchair GNR is pulled along the armchair direction of the substrate (see Figure 1). For nanoscale interfaces one often finds a typical scaling of the friction force with the contact size ranging from linear in commensurate contacts to sublinear in disordered and incommensurate ones.[20, 48-51] As is clearly evident in Figure 2a-b, the aligned motion of GNRs atop graphene or $h$-BN surfaces exhibits a qualitatively different behavior. Both aligned junctions display an initial linear increase of the static and kinetic friction forces with the ribbon length ($L_{\mathrm{GNR}}$) that is followed by leveling-off above a characteristic length of $L_{\mathrm{GNR}}$~10 nm and ~20 nm for the homogeneous and heterogeneous interfaces, respectively.

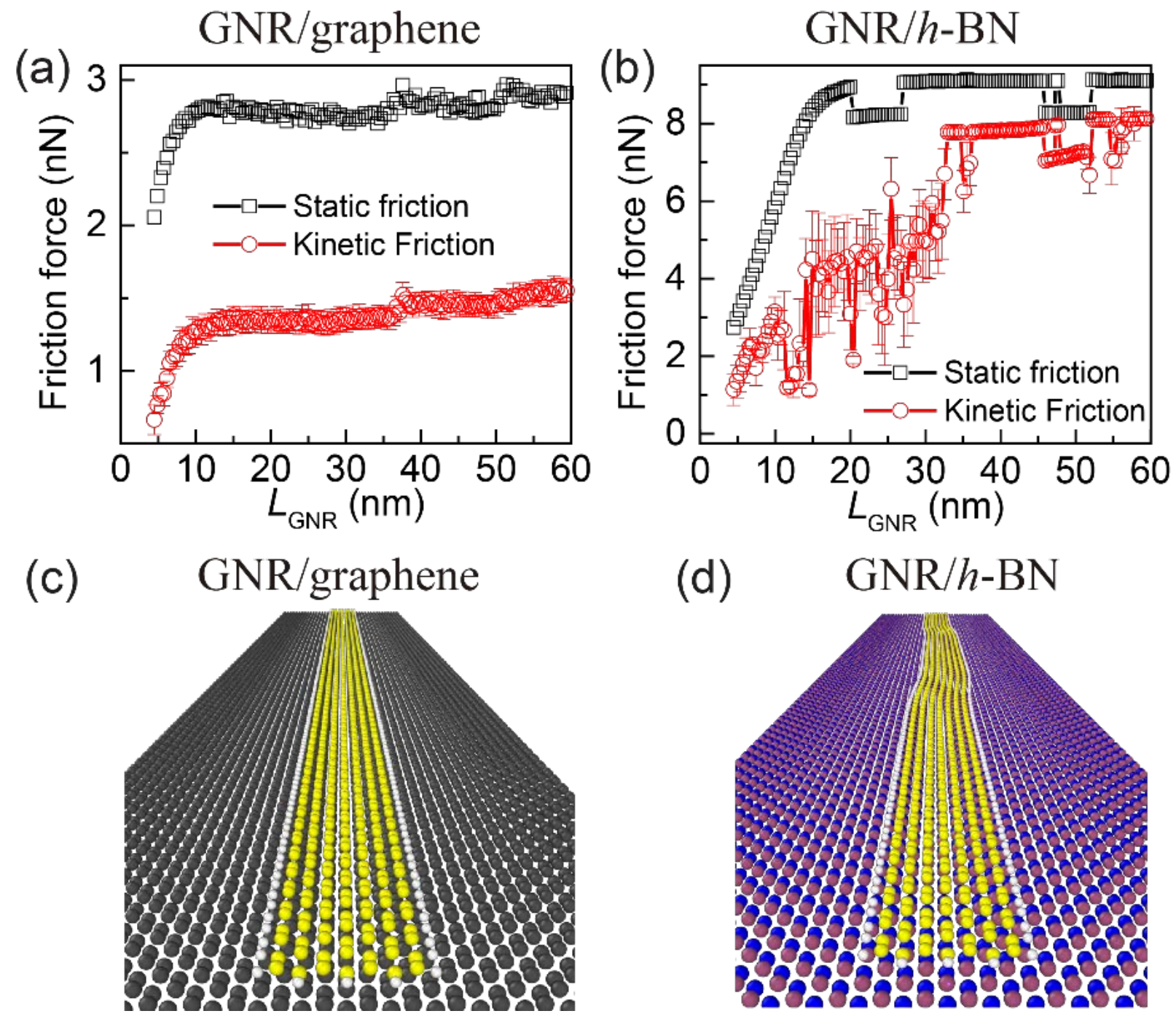

***Figure 2.*** *Static (black rectangles) and kinetic (red circles) friction of GNRs sliding over (a) graphene and (b) h-BN as functions of their length. The optimized geometries of a 36.76 nm long armchair GNR deposited along the armchair axis*

*of graphite and h-BN substrates are presented in panels (c) and (d), respectively. Mauve, blue, yellow, and grey spheres represent boron, nitrogen, carbon, and hydrogen atoms, respectively.*

Another counterintuitive behavior that is demonstrated in Figure 2a,b is the fact that at almost any given GNR's length, the friction forces of the heterogeneous contact are found to be ~3-fold larger than those of its homogeneous counterpart. This is in striking contrast with the commonly accepted paradigm that incommensurate interfaces between rigid layered materials should exhibit lower friction than the corresponding homogeneous ones.[8, 15, 20, 52-55] Furthermore, while the homogeneous contact exhibits a smooth variation of the static and kinetic frictional forces with the GNR's length, the heterojunction shows a much richer behavior, characterized by strong fluctuations of the kinetic friction and sudden jumps of the static friction.

Notably, both aligned homogeneous and heterogeneous junctions also display significantly different relaxed configurations and modes of motion as a function of their length. Upon geometry optimization, all GNRs deposited on a graphene substrate obtain straight configurations (see Figure 2c). On the contrary, when deposited atop of an *h*-BN substrate, GNRs of length 20.4 nm and beyond exhibit a buckled structure (see Figure 2d). These initial relaxed configurations may be dynamically modified during sliding. When pulled along a graphene substrate, short GNRs keep their straight configuration with negligible structural deformations in the lateral direction perpendicular to the sliding direction (see Supplementary Movie 1). In contrast, beyond a length of ~25 nm, the GNRs exhibit shear-induced buckling that results in snake-like motion (see Supplementary Movie 2). A completely different picture arises for GNRs sliding atop an *h*-BN substrate, where short ribbons exhibit in-plane bending and irregular motion (see Supplementary Movie 3) whereas ribbons of length 35.5 nm and beyond experience shear-induced unbuckling followed by nearly unidirectional motion (see Supplementary Movie 4).

To rationalize these intriguing findings we first analyze the stress distribution along the GNR main axis during the pulling process and its effect on the length-dependence of the frictional forces. Panels (a) and (b) of Figure 3 illustrate the stress distribution along the GNR as a function of pulling time. Focusing first on the homogeneous junction (Figure 3a) the motion is characterized by stick-slip dynamics. Upon pulling, stress nucleation occurs, growing from the leading edge into the GNR bulk. This is followed by an abrupt stress propagation towards the trailing edge, resulting in a global slip event. We note here that depending on the local stacking mode, stress may also develop near the center of the GNR. Considering the stress distribution at the onset of sliding we find that near the pulling edge it can be well fitted with an exponential function (Figure 3c). Hence, we can assess the stress penetration depth to be $L_c \sim 4.14$ nm, which is considerably shorter than the overall ribbon length of 27.5 nm. The former, is dictated by the ratio between the in-plane GNR stiffness ($K_{\mathrm{GNR}}$) and the ribbon/substrate interaction stiffness ($K_{\mathrm{Interface}}$) via $L_{\mathrm{c}} = L_{\mathrm{GNR}}\sqrt{K_{\mathrm{GNR}}/K_{\mathrm{Interface}}}$ (see Section 8 of the SI for further details).[56]

This observation provides an explanation for the variation of the friction force with ribbon length. For GNRs shorter than the characteristic stress penetration depth a linear increase of both static and kinetic friction forces is obtained, as expected for commensurate junctions. Once the ribbon length exceeds the stress penetration depth only the atoms in the vicinity of the pulling edge experience stretching and the rest of the bulk atoms remain in their relaxed configuration until the sliding event occurs. Hence, the elastic energy stored during the nucleation stage becomes independent of the ribbon's length, resulting in friction forces leveling-off. We note that the residual increase of kinetic friction observed above $L_{\mathrm{GNR}}$=40 nm (see red circles in Figure 2a) is caused by the contribution of the viscous-like dissipation term in eq 1, which is proportional to the number of atoms to which damping is applied.

A qualitatively similar stick-slip behavior is also found for the heterojunction (Figure 3b). However, in this case a much broader non-exponential stress distribution (Figure 3d) is obtained during the nucleation stage. This difference is clearly demonstrated in panels (e) and (f) of Figure 3, where the per-atom stress distribution at the onset of sliding of the heterojunction penetrates much deeper into the GNR bulk than for its homogeneous counterpart. Therefore, the leveling-off of the friction forces occurs at a considerably longer GNR length of ~20 nm (see Figure 2b). We note that the hydrogen passivated edge atoms possess a slightly shorter C-C equilibrium bond distance than their bulk counterparts as indicated by their blue coloring in panels (e) and (f) of Figure 3.

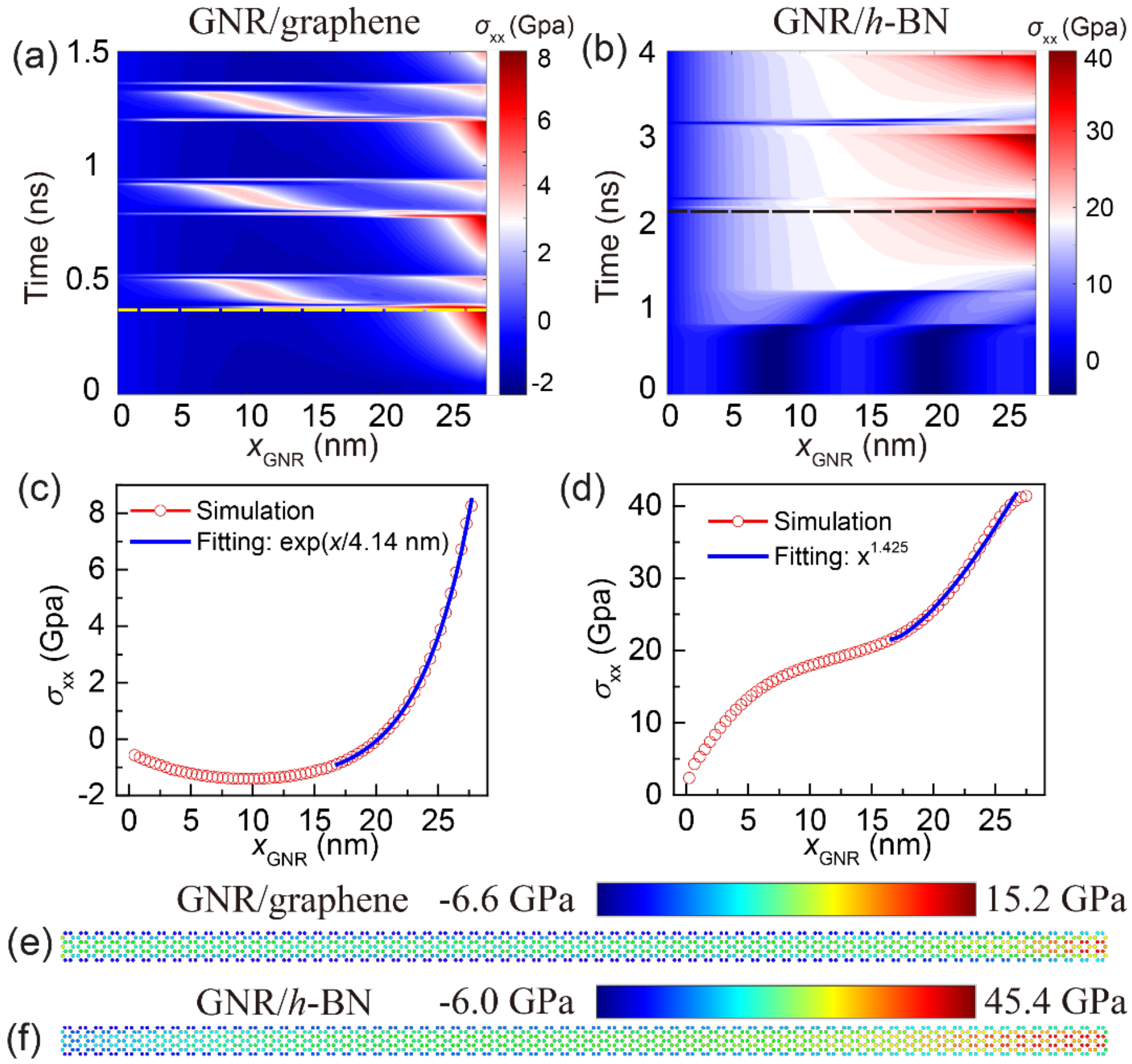

***Figure 3.*** *Stress distribution within a 28 nm long GNR sliding atop graphene (left) and h-BN (right) substrates. Panels (a) and (b) present time evolution 2D color maps of the stress distribution along the GNR, for the homogeneous (a) and heterogeneous (b) junctions. Panels (c) and (d) show the cross section of the above 2D maps at times that correspond to the orange and black horizontal dashed lines of panels (a) and (b), respectively, at the onset of global motion. All stresses reported are calculated by averaging the per-atom stresses along the narrow dimension of the GNR (excluding the passivating hydrogen atoms) and over a single axial unit-cell. The position of each axial unit-cell along the GNR is calculated as the distance of its center-of-mass from the ribbon's trailing edge. Panels (e) and (f) show the per-atom stress distributions along the GNR that correspond to the onset of sliding of the GNR on graphene and h-BN substrates, respectively (passivating hydrogen atoms are not shown). Note the different scale that the color bars represent in the two panels.*

To explain the different relaxed configurations and the irregular behavior of the friction force with ribbon length in the heterojunction (Figure 2b), interfacial commensurability and the formation of moiré superstructures must be taken into account. In the case of extended interfaces, the inherent 1.8% mismatch between graphene and *h*-BN lattice vectors is locally compensated via in-plane deformations. The mating layers form regions of nearly perfect registry and interlayer distance that are separated by elevated ridges to partially alleviate the ensuing strain by exploiting the softer out-of-plane bending modes.[20, 57, 58] For aligned contacts, the periodicity of these moiré patterns is of $L_{\text{moiré}}$~14 nm.[18] In the (quasi)one-dimensional case of the GNR, energy minimization can be achieved not only via in-plane compression and stretching and out-of-plane displacements but also via lateral buckling in the direction perpendicular to the main ribbon axis. For short GNRs (below 20.4 nm for the ribbon width considered herein) the energy cost of such buckling is too high and the ribbon preserves its straight geometry. As the length of the ribbon increases, the competition between the intra-layer (elastic) energy and the quest for interlayer registry matching results in the onset of buckling (see Figure 4d,f).

The shear-induced dynamics of these undulations, which share the periodicity of the moiré pattern, are manifested in the length dependence of the static and kinetic friction exhibited by the heterogeneous junction. After reaching the plateau, the static friction shows sharp jumps between two distinct values (~9.1 nN and ~8.3 nN) with increasing ribbon length. The higher (lower) static friction values correspond to GNRs exhibiting even (odd) number of buckles, where the leading edge of the ribbon is positioned in an energetically (un)favorable stacking mode (see corresponding snapshots presented in SI Section 9). During sliding, the interplay between in-plane ribbon elasticity and its interaction with the *h*-BN substrate, leads to complex dynamics involving ribbon bending and irregular motion for the shorter GNRs (see Supplementary Movie 3). This dynamics is responsible for the erratic length dependence of the kinetic friction force exhibited by the shorter GNRs (see Figure 2b). A more regular length dependence of the kinetic friction is found for the longer GNRs that exhibit nearly unidirectional steady-state motion (see Supplementary Movie 4).

The origin of the higher friction force exhibited by the heterojunctions with respect to their homogeneous counterparts lies in the difference of the energy barriers encountered during the sliding motion. Previously, we found a similar effect for small two-dimensional graphene flakes sliding atop graphene and *h*-BN surfaces.[20] There, when moving atop graphene, the center of mass of the graphene flake slider performed zig-zag type of motion and avoided the global energy barriers resulting in a less corrugated energy path. The sliding energy surface of the heterojunction possessed a more corrugated minimal energy path thus leading to higher frictional forces. Similar phenomena are obtained for the leading edge of the GNRs studied herein. The GNR head in the homogeneous junction performs noticeable zigzag motion (see Supplementary Movies 2) to reduce the sliding potential energy barriers along its path and hence reduce the friction. Interestingly, the adaptation of the GNR backbone to its head's rattling translates into the snake-like motion discussed above. In the heterogeneous case, following the initial unbuckling stage, the GNR's head deviates much less from the sliding axis. Hence, its body performs nearly unidirectional motion characterized by a more corrugated sliding energy path and dissipative stick-slip motion (see steady-states in Figure 4a and b and Supplementary Movies 4).

Additional information regarding the shear-induced dynamics of the GNRs can be obtained by further analyzing the friction trace. The aligned homogeneous junctions (see Figure 4c and e) show a very regular double-peaked stick-slip behavior (see Figure 4a). The difference in height of the two peaks reflects the fact that the onset of slip motion results from a pullout of the leading edge atoms from potential energy surface minima of different depth along the sliding path (see Supporting Movie 5). The force traces of the aligned heterogeneous junctions are quite different exhibiting a sequence of precursor partial slips prior to the onset of global sliding (see Figure 4b). These events reflect the progressive straightening of the ribbons that occurs via successive elimination of the buckled regions (see Figure 4d and f and SI Movie 4), starting from the leading edge and advancing towards the end. Upon complete straightening of the GNR, a global slip event takes place (see Figure 4b). We note that similar force traces, exhibiting partial slip events preceding global sliding, have been observed in macroscale experiments. [59, 60] Nevertheless, their origin lies in the evolution of contact area rather than shear-induced unbuckling.

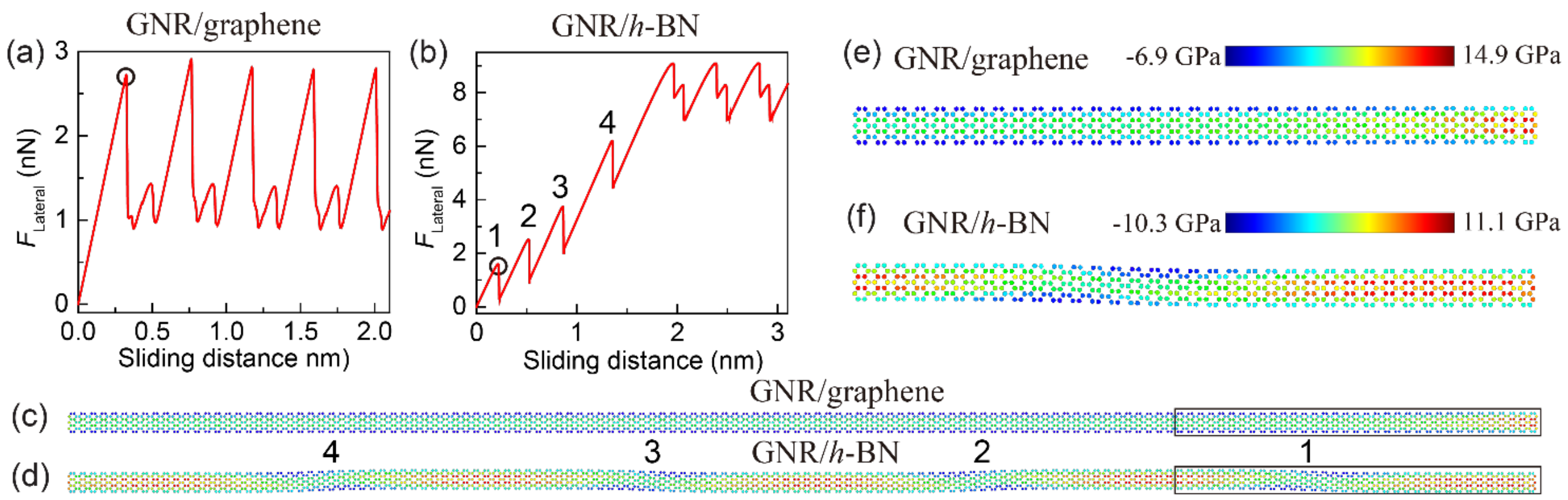

***Figure 4.*** *Pulling force as a function of the sliding distance of a GNR deposited over (a) graphene and (b) h-BN substrates. (c)-(d) The corresponding stress distributions of a GNR of length 59.4 nm, computed at the onset of the first (partial) slip event, as indicated by the circles in panels (a) and (b). Panels (e) and (f) provide zoom-in on the rectangular regions highlighted in panels (c) and (d), respectively. Passivating hydrogen atoms are not shown.*

## Misaligned Interfaces

To study the friction dependence on the misfit angle between the ribbon and the surface we performed similar simulations while pulling the GNRs at two angles of 45° and 90° with respect to the armchair axis of the substrate. As may be expected due to incommensurability considerations, the latter (not shown) exhibit smooth sliding accompanied by ultra-low friction regardless of the underlying surface.[8, 11, 13, 61, 62] Pulling at the angle of 45° results in a more diverse behavior, as illustrated in Figure 5. Panels (a) and (c) present the lateral force traces (red curves) for the homogeneous GNR/graphene junction with ribbon lengths of 4.5 nm and 27.5 nm, respectively. After a short interval characterized by smooth sliding, a sudden increase of friction occurs, reflecting an abrupt reorientation of the ribbon to achieve an energetically more favorable interfacial registry with the underlying surface. As shown by the blue lines in panels (a) and (c) of Figure 5 the ribbon rotates from its original $45^{\circ}$ alignment to an average angle of $\sim 60^{\circ}$ with respect to the armchair direction of the graphene substrate (see also Supplementary Movie 6). Comparing panels (a) and (c) we find that the shorter the GNR is, the earlier its reorientation occurs during the dynamics. We note that similar reorientation processes have been observed experimentally and computationally for graphene flakes sliding atop a graphite surface.[63, 64]

The shorter heterogeneous GNR/*h*-BN junction (Figure 5b) exhibits a very similar behavior to that of its homogeneous counterpart with an initial low-friction stage followed by a rotation towards the $60^{\circ}$ misaligned configuration that is accompanied by a sharp increase of friction (see Figure 5b,f). On the contrary, the frictional dynamics of the longer heterojunction is characterized by a gradual reorientation process (Figure 5d). This results from the shear-induced dynamics of the ribbon's buckled structure leading to snake-like motion (see Supplementary Movie 7).

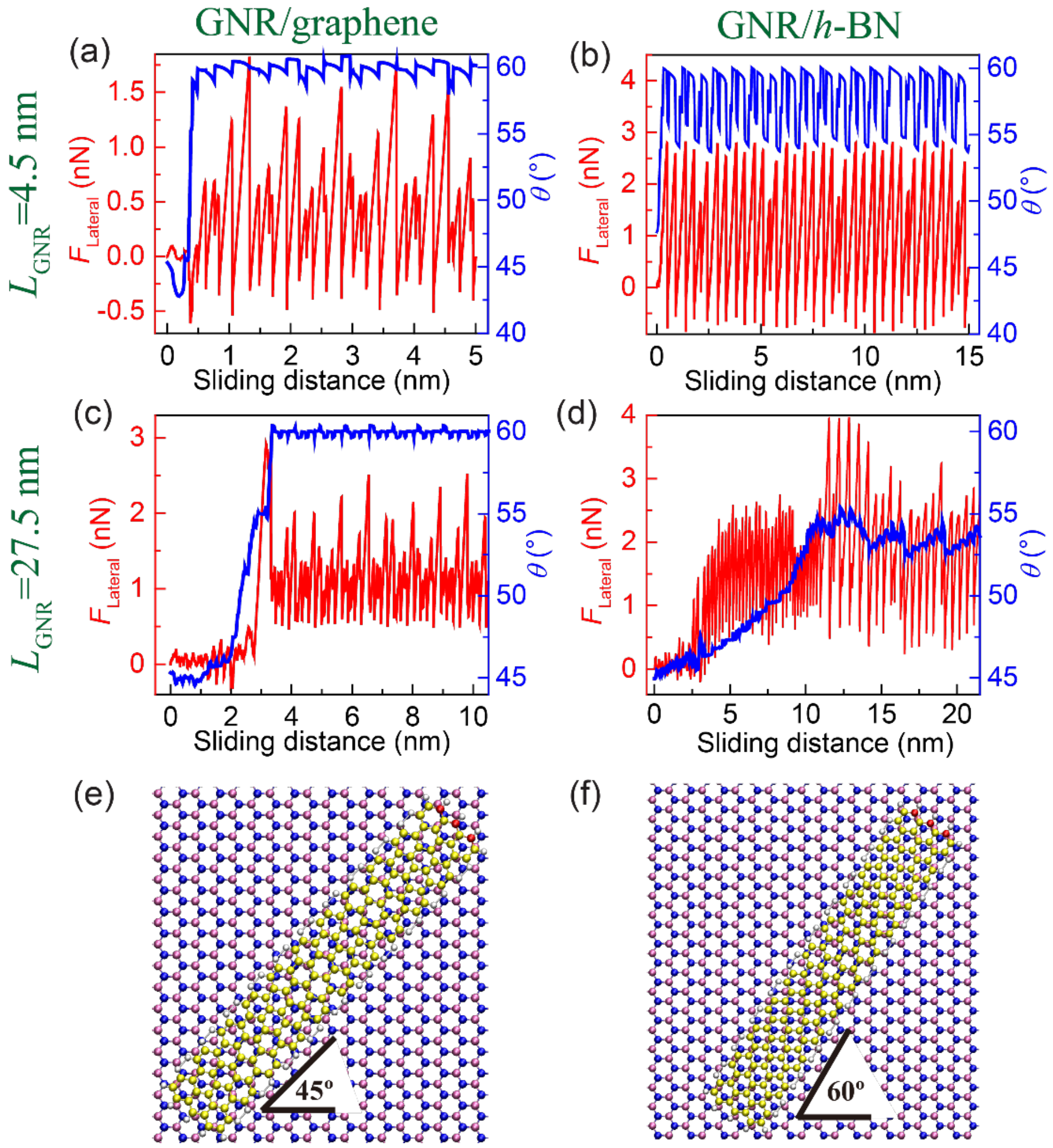

***Figure 5.*** *Frictional motion of GNRs pulled along a direction of* $45^{\circ}$ *with respect to the armchair direction of graphene ((a) and (c)) and h-BN ((b) and (d)) substrates. Both the lateral force (left axis, red) and the average angle (right axis, blue) are presented as a function of sliding distance for* $L_{GNR} = 4.5\ nm$ *(panels (a) and (b)) and* $27.5\ nm$ *(panels (c) and (d)). The configurations of the shorter GNR on h-BN before and after reorientation are presented in panels (e) and (f), respectively. Mauve, blue, yellow, and grey spheres represent boron, nitrogen, carbon, and hydrogen atoms, respectively.*

## Summary and Conclusions

Quasi-one-dimensional junctions between layered materials introduce an additional degree of freedom over their extended two-dimensional counterparts. Apart from in-plane compression/expansion and out-of-plane corrugation, nanoribbons are allowed to buckle in the direction perpendicular to the sliding motion in order to enhance their registry with the underlying substrate. This leads to new types of driven motion and frictional behavior. Specifically, both homogeneous and heterogeneous junctions of a GNR aligned with graphene and *h*-BN exhibit a length independence of the friction beyond a certain contact size due to the finite penetration of the in-plane stress into the GNR bulk. In the homogeneous case, in order to follow a less corrugated energy path, the ribbon's head performs a zig-zag movement, which translates to less dissipative snake like motion of the longer systems. On the contrary, the longer heterojunctions exhibit initial shear-induced unbuckling followed by nearly unidirectional dissipative stick-slip motion. The interplay between the stress distribution along the GNR and the lattice mismatch of the contacting surfaces dictates the characteristic length scales at which these phenomena will take place. Misaligned contacts show either ultra-low friction or dynamic transition between low and high friction states due to shear induced reorientations. The simulation of such intricate phenomena became possible by a LAMMPS implementation of our registry-dependent interlayer potential and refinement of its parameterization for bilayer graphene and its heterojunction with *h*-BN.

**Supporting Information**.

The supporting Information section includes a description of: the ILP parameters refinement procedure; benchmark tests for the ILP LAMMPS implementation; sensitivity tests of the results with respect to the ILP parameterization, choice of intralayer potential, values of the damping coefficients, propagation time step, and applied temperature; theoretical estimation of the characteristic stress decay length into the GNR bulk, and stacking modes of the GNR's leading edge atop an *h*-BN substrate.

**Corresponding Author**

* E-mail: urbakh@post.tau.ac.il.

**Notes**

The authors declare no competing financial interest.

**Acknowledgments**

We would like to thank Prof. Leeor Kronik and Dr. Ido Azuri for their help with the reference calculations and for helpful discussions. W. O. acknowledges the financial support from a fellowship program for outstanding postdoctoral researchers from China and India in Israeli Universities. D. M. acknowledges the fellowship from the Sackler Center for Computational Molecular and Materials Science at Tel Aviv University, and from Tel Aviv University Center for Nanoscience and Nanotechnology. M. U. acknowledges the financial support of the Israel Science Foundation, Grant No.1316/13, and of the Deutsche Forschungsgemeinschaft (DFG), Grant No. BA 1008/21-1. O. H. is grateful for the generous financial support of the Israel Science Foundation under grant no. 1586/17 and the Naomi Foundation for generous financial support via the 2017 Kadar Award. This work is supported in part by COST Action MP1303.

# Nanoserpents: Graphene Nanoribbons Motion on Two-Dimensional Hexagonal Materials

# Supporting Information

Wengen Ouyang, Davide Mandelli, Michael Urbakh,* Oded Hod

*School of Chemistry, and The Sackler Center for Computational Molecular and Materials Science, Tel Aviv University, Tel Aviv 6997801, Israel.*

**Corresponding Author**

* E-mail: urbakh@post.tau.ac.il

This supporting information document includes the following sections:

## 1. <u>Refined Fitting Parameters of the Registry Dependent Interlayer Potential for Graphene and *h*-BN</u>

The registry dependent interlayer potential (ILP) and the Kolmogorov Crespi (KC) potential have the following general pairwise form:[1-3]

$$V(\mathbf{r}_{ij},\mathbf{n}_i,\mathbf{n}_j) = \mathrm{Tap}(r_{ij})\left[V_{\mathrm{att}}(r_{ij}) + V_{\mathrm{Rep}}(\mathbf{r}_{ij},\mathbf{n}_i,\mathbf{n}_j) + V_{\mathrm{Coul}}(r_{ij})\right]. \quad (1)$$

Here, $V_{\mathrm{att}}(r_{ij})$, $V_{\mathrm{Rep}}(\mathbf{r}_{ij},\mathbf{n}_i,\mathbf{n}_j)$, and $V_{\mathrm{Coul}}(r_{ij})$ correspond to the long-range van der Waals attraction, short-range Pauli repulsion, and monopolar electrostatic interactions, respectively. These terms take different forms in the KC and ILP potentials as detailed below. $\mathbf{r}_{ij}$ is the vector distance between atoms $i$ and $j$ residing on different layers, while $\mathbf{n}_k$ is a unit vector normal to the surface at the $k^{\mathrm{th}}$ atomic position. The latter is defined as the average of the three vectors normal to the planes defined by the triangles formed by the $k^{\mathrm{th}}$ atom with its three nearest neighbors within the hexagonal lattice. These three normals are calculated as the cross products between the displacement vectors from atomic position $k$ to two of its nearest neighbors, considering each distinct couple of nearest neighbors.[4] In open boundary systems, the atoms at the edges have only one or two nearest neighbors. The normal to an atom having two nearest neighbors is calculated as the cross product between the displacement vectors to its two nearest neighbors. In the case of an atom that has only one nearest neighbor, first the cutoff is adjusted in order to include one or two second nearest neighbors; the normal is then computed following the appropriate procedure out of the two outlined above. The taper function

$$\mathrm{Tap}(r_{ij}) = 20\left(\frac{r_{ij}}{R_{\mathrm{cut},ij}}\right)^7 - 70\left(\frac{r_{ij}}{R_{\mathrm{cut},ij}}\right)^6 + 84\left(\frac{r_{ij}}{R_{\mathrm{cut},ij}}\right)^5 - 35\left(\frac{r_{ij}}{R_{\mathrm{cut},ij}}\right)^4 + 1 \quad (2)$$

provides a continuous long-range cutoff (up to third derivative) that dampens the various interactions at interatomic separations larger than $R_{\mathrm{cut},ij}$.

### 1.1 The Interlayer Potential (ILP)

The analytical form of the long-range attractive term is adapted from the Tkatchenko-Scheffler augmentation scheme[5] to density functional theory (DFT) given by the standard $r^{-6}$ expression dampened at short range by a Fermi-Dirac type function, which in DFT calculations avoids double counting of interactions:

$$V_{\mathrm{att}}(r_{ij}) = -\frac{1}{1+e^{-d_{ij}\left[r_{ij}/\left(s_{R,ij}\cdot r_{ij}^{\mathrm{eff}}\right)-1\right]}}\frac{C_{6,ij}}{r_{ij}^{6}}. \tag{3}$$

Here, $C_{6,ij}$ is the pairwise dispersion coefficient of atoms $i$ and $j$ residing on adjacent layers, $r_{ij}^{\mathrm{eff}}$ is the sum of their effective equilibrium vdW atomic radii, and $d_{ij}$ and $s_{R,ij}$ are unit-less parameters defining the steepness and onset of the short-range Fermi−Dirac type damping function.

The repulsive term is written as a combination of isotropic and anisotropic contributions as follows:

$$V_{\mathrm{Rep}}(\mathbf{r}_{ij},\mathbf{n}_i,\mathbf{n}_j) = e^{\alpha_{ij}\left(1-\frac{r_{ij}}{\beta_{ij}}\right)}\left\{\varepsilon_{ij} + C_{ij}\left[e^{-\left(\rho_{ij}/\gamma_{ij}\right)^2} + e^{-\left(\rho_{ji}/\gamma_{ij}\right)^2}\right]\right\}, \tag{4}$$

where $\varepsilon_{ij}$ and $C_{ij}$ are constants that set the energy scales associated with the isotropic and anisotropic repulsion, respectively, $\beta_{ij}$ and $\gamma_{ij}$ set the corresponding interaction ranges, and $\alpha_{ij}$ is a parameter that sets the steepness of the isotropic repulsion function. The lateral interatomic distance $\rho_{ij}$ is defined as the shortest distance from atom $j$ to the surface normal, $\mathbf{n}_i$, at the position of atom $i$:

$$\begin{cases}\rho_{ij}^2 = r_{ij}^2 - \left(\mathbf{r}_{ij}\cdot\mathbf{n}_i\right)^2 \\ \rho_{ji}^2 = r_{ji}^2 - \left(\mathbf{r}_{ji}\cdot\mathbf{n}_j\right)^2\end{cases}. \tag{5}$$

The electrostatic term, which appears only in the homogeneous $h$-BN ILP, is given by a shielded monopolar Coulomb expression of the form:

$$V_{\mathrm{Coul}}(r_{ij}) = kq_iq_j/\sqrt[3]{r_{ij}^3 + \lambda_{ij}^{-3}}. \tag{6}$$

Here, $k = 14.399645\ \mathrm{eV}\cdot\mathrm{Å}\cdot\mathrm{C}^{-2}$ is Coulomb's constant, while $q_i$ and $q_j$ are the effective charges of atoms $i$ and $j$ (residing in different layers) given in units of the absolute value of the electron charge, $e$, and $\lambda_{ij} = \sqrt{\lambda_{ii}\lambda_{jj}}$ is a shielding parameter used to eliminate the short-range singularity of the electrostatic interaction in regions where the Pauli repulsions between overlapping electron clouds dominate the interlayer potential. In the present study, we used the fixed effective atomic charge approximation adopting values of $q_B = 0.42e$ and $q_N = -0.42e$.[1]

## 1.2 The Kolmogorov Crespi Potential

The van de Waals attraction term of the KC potential has the following form[4]:

$$V_{\mathrm{att}}(r_{ij}) = -A_{ij}\left(\frac{z_{0,ij}}{r_{ij}}\right)^6, \tag{7}$$

where $A_{ij}$ and $z_{0,ij}$ are energy and length scale parameters, respectively. The anisotropic repulsion term reads:

$$V_{\mathrm{Rep}}\left(\mathbf{r}_{ij},\mathbf{n}_i,\mathbf{n}_j\right)=e^{-\lambda_{ij}\left(r_{ij}-z_{0,ij}\right)}\left\{C_{ij}+e^{-\left(\rho_{ij}/\delta_{ij}\right)^2}\left[C_{0,ij}+C_{2,ij}\left(\frac{\rho_{ij}}{\delta_{ij}}\right)^2+C_{4,ij}\left(\frac{\rho_{ij}}{\delta_{ij}}\right)^4\right]+e^{-\left(\rho_{ji}/\delta_{ij}\right)^2}\left[C_{0,ij}+C_{2,ij}\left(\frac{\rho_{ji}}{\delta_{ij}}\right)^2+C_{4,ij}\left(\frac{\rho_{ji}}{\delta_{ij}}\right)^4\right]\right\}, \quad (8)$$

where $C_{ij}$, $\lambda_{ij}$ and $C_{0/2/4,ij}$, $\delta_{ij}$ are energy and length scale parameters of the isotropic and anisotropic repulsion terms, respectively. In the KC potential, monopolar electrostatic interactions are neglected and no long-range cut-off is applied.

### 1.3 Fitting procedure

In the expressions presented above, the ILP parameters $\alpha_{ij},\beta_{ij},\gamma_{ij},\varepsilon_{ij},C_{ij},d_{ij},s_{R,ij},r_{ij}^{\mathrm{eff}},C_{6,ij},R_{\mathrm{cut},ij},\lambda_{ij}$ and the KC parameters $z_{0,ij},A_{ij},\delta_{ij},C_{ij},C_{0,ij},C_{2,ij},C_{4,ij},\lambda_{ij}$ serve as fitting parameters. Here, we provide two refined sets of parameters for the registry dependent ILP for homogeneous interfaces of graphene and hexagonal boron nitride (*h*-BN), as well as their heterojunctions and one set of refined parameters for the KC potential for graphene based systems. The force-field has been benchmarked against density functional theory calculations of several dimer systems within the Heyd-Scuseria-Ernzerhof hybrid density functional approximation,[6-8] corrected for many-body dispersion effects (see section S2 below).[9, 10] Unlike the previous parametrizations,[1, 2] where the parameters were fitted manually focusing on achieving good agreement only in the long-range interaction regime, in the present parametrization the parameters were fitted using an automatic interior-point technique, as implemented in MATLAB,[11, 12] which improved the agreement with the reference DFT data across the entire interaction region.

Our training set included three periodic structures (graphene/graphene, graphene/*h*-BN and *h*-BN/*h*-BN) and 10 finite structures (Benzene dimer, Borazine dimer, $B_{12}N_{12}H_{12}$ dimer, Coronene dimer, Benzene/Coronene, Borazine/$B_{12}N_{12}H_{12}$, Benzene/Borazine, Benzene/$B_{12}N_{12}H_{12}$, Borazine/Coronene and Coronene/$B_{12}N_{12}H_{12}$). The reference data consisted of binding energy curves (Figure S1-S3) and sliding energy surfaces (Figure S4-S5) of all systems. The latter were computed fixing the equilibrium interlayer distance to that of the optimal stacking mode of the corresponding periodic structures. For the case of heterogeneous graphene/*h*-BN junctions we considered two binding energy curves calculated at the optimal

(C-) and worst (A-) stacking modes.[2]

The fitting procedure involved two steps. First, we fitted the parameters for the three periodic structures, using both the binding energy curves and the sliding potential surfaces. This provided us with the C-C, B-B, N-N, C-B, C-N, and B-N sets of parameters. Next, we fixed these parameters and fitted the remaining H-H, C-H, B-H, and N-H parameter sets using the reference data corresponding to the finite dimers. In this final stage we introduced a weighting factor proportional to the dimer size to increase the importance of the larger dimers during the fitting procedure. The resulting ILP parameters are presented in Table S1.

**Table S1**: List of ILP parameter values for graphene and *h*-BN based systems. The training set includes all the binding energy curves and all the sliding potential surfaces mentioned in the text. A value of $R_{\mathrm{cut}} = 16$ Å is used throughout.

| | $\beta_{ij}$ (Å) | $\alpha_{ij}$ | $\gamma_{ij}$ (Å) | $\varepsilon_{ij}$ (meV) | $C_{ij}$ (meV) | $d_{ij}$ | $s_{R,ij}$ | $r_{eff,ij}$ (Å) | $C_{6,ij}$ (eV•Å$^6$) | $\lambda_{ij}$ (Å$^{-1}$) |
|---|---|---|---|---|---|---|---|---|---|---|
| **C-C** | 3.2058 | 7.5111 | 1.2353 | 1.53E-05 | 37.5304 | 15.4999 | 0.7954 | 3.6814 | 25.7145 | -- |
| **B-B** | 3.1437 | 9.8251 | 1.9364 | 2.7848 | 14.4960 | 15.1993 | 0.7834 | 3.6829 | 49.4980 | 0.70 |
| **N-N** | 3.4432 | 7.0845 | 1.7473 | 2.9140 | 46.5086 | 15.0204 | 0.8008 | 3.5518 | 14.8102 | 0.69 |
| **H-H** | 3.9745 | 6.5380 | 1.0806 | 0.6701 | 0.8334 | 15.0224 | 0.7491 | 2.7672 | 1.6160 | -- |
| **C-B** | 3.3037 | 10.5441 | 2.9267 | 16.7200 | 0.3572 | 15.3053 | 0.7002 | 3.0973 | 30.1629 | -- |
| **C-N** | 3.2536 | 8.8259 | 1.0595 | 18.3447 | 21.9136 | 15.0000 | 0.7235 | 3.0131 | 19.0631 | -- |
| **B-N** | 3.2953 | 7.2243 | 2.8727 | 1.3715 | 0.4347 | 14.5946 | 0.8044 | 3.7657 | 24.6700 | 0.694982 |
| **C-H** | 2.6429 | 12.9141 | 1.0203 | 0.9750 | 25.3410 | 15.2229 | 0.8116 | 3.8873 | 5.6875 | -- |
| **B-H** | 2.7187 | 9.2146 | 3.2731 | 14.0157 | 14.7605 | 15.0848 | 0.7768 | 3.6409 | 7.9642 | -- |
| **N-H** | 2.7535 | 8.2267 | 3.1064 | 0.8074 | 0.3944 | 15.0332 | 0.7451 | 2.7336 | 3.8462 | -- |

The training set for the parameters presented in Table S1 included the binding energy curve of the energetically least favorable A-stacked graphene/*h*-BN junction. As a consequence, for the heterojunction we observe a somewhat larger deviation of the ILP results from the reference sliding energy potential compared to that obtained in the homogeneous cases (see Figure S4 and S5). In Table S2 we present a second set of ILP parameters that was obtained excluding the A-stacked graphene/$h$-BN binding energy curve from the training set, which improves the agreement with the reference DFT data. Specifically, for commensurate heterojunctions we suggest using Table S2 parameters when calculating tribological properties at the equilibrium interlayer distance, whereas Table S1 parameters should be used for calculations in the sub-equilibrium regime. For incommensurate graphene/$h$-BN heterojunctions the two parameter sets provide similar results at equilibrium interlayer distance (see Figure S7) and can be both used.

**Table S2**: List of ILP parameter values for graphene and $h$-BN based systems. The training set is the same as that of Table S1 apart for the exclusion of the binding energy curve calculated at the A-stacking mode of the graphene/$h$-BN junction. A value of $R_{\mathrm{cut}} = 16$ Å is used throughout.

| | $\beta_{ij}$ (Å) | $\alpha_{ij}$ | $\gamma_{ij}$ (Å) | $\varepsilon_{ij}$ (meV) | $C_{ij}$ (meV) | $d_{ij}$ | $s_{R,ij}$ | $r_{eff,ij}$ (Å) | $C_{6,ij}$ (eV•Å$^6$) | $\lambda_{ij}$ (Å$^{-1}$) |
|---|---|---|---|---|---|---|---|---|---|---|
| **C-C** | 3.2058 | 7.5111 | 1.2353 | 1.53E-05 | 37.5304 | 15.4999 | 0.7954 | 3.6814 | 25.7145 | -- |
| **B-B** | 3.1437 | 9.8251 | 1.9364 | 2.7848 | 14.4960 | 15.1993 | 0.7834 | 3.6829 | 49.4980 | 0.70 |
| **N-N** | 3.4432 | 7.0845 | 1.7473 | 2.9140 | 46.5086 | 15.0204 | 0.8008 | 3.5518 | 14.8102 | 0.69 |
| **H-H** | 3.4994 | 6.5011 | 1.4887 | 0.0044 | 2.1538 | 15.2527 | 0.7090 | 2.6454 | 1.3485 | -- |
| **C-B** | 3.0957 | 11.4129 | 3.5402 | 0.0067 | 0.0021 | 15.4960 | 0.7727 | 3.3415 | 31.1639 | -- |
| **C-N** | 3.2371 | 8.3963 | 1.5489 | 18.2309 | 31.8545 | 15.0000 | 0.8100 | 3.7858 | 18.8623 | -- |
| **B-N** | 3.2953 | 7.2243 | 2.8727 | 1.3715 | 0.4347 | 14.5946 | 0.8044 | 3.7657 | 24.6700 | 0.694982 |
| **C-H** | 2.6478 | 10.7335 | 5.9574 | 37.2437 | 0.7124 | 15.2182 | 0.7126 | 2.6665 | 5.8883 | -- |
| **B-H** | 2.6498 | 9.8478 | 2.9422 | 0.3973 | 22.1276 | 15.4635 | 0.8498 | 3.4991 | 6.4569 | -- |
| **N-H** | 2.8599 | 8.5956 | 5.6698 | 0.0080 | 0.0039 | 15.1037 | 0.8499 | 3.4995 | 3.1446 | -- |

The corresponding set of refined KC potential parameters is given in Table S3.

**Table S3**: List of KC parameter values for graphene based systems. The training set includes all the binding energy curves of graphene-based systems and the sliding potential surface of periodic bilayer graphene.

| | $z_{0,ij}$ (Å) | $C_{0,ij}$ (meV) | $C_{2,ij}$ (meV) | $C_{4,ij}$ (meV) | $C_{ij}$ (meV) | $\delta_{ij}$ (Å) | $\lambda_{ij}$ (Å$^{-1}$) | $A_{ij}$ (meV) |
|---|---|---|---|---|---|---|---|---|
| **C-C** | 3.3288 | 21.8472 | 12.0602 | 4.7111 | 6.6789E-04 | 0.77181 | 3.1439 | 12.6603 |
| **C-H** | 3.1565 | 37.4005 | 8.3911E-03 | 55.0618 | 5.18E-05 | 0.44373 | 2.5088 | 11.4791 |
| **H-H** | 2.2188 | 4.53E-05 | 4.87E-05 | 2.02774146 | 1.19395 | 0.89685 | 0.238105 | 9.22E-05 |

## 2. <u>Implementation of the ILP and KC Potentials within the LAMMPS Package and Benchmark Tests</u>

We have implemented the ILP and KC potential within the LAMMPS package for molecular dynamics simulations.[13] In the next sections, we report the results of a set of benchmark calculations used to check the agreement between our implementation of the ILP and KC potential and the reference DFT data.

### 2.1 Binding Energy Curves

Figure S1 presents the binding energy curves calculated for the laterally periodic bilayer structures, using the two sets of parameters reported in Table S1 and S2.

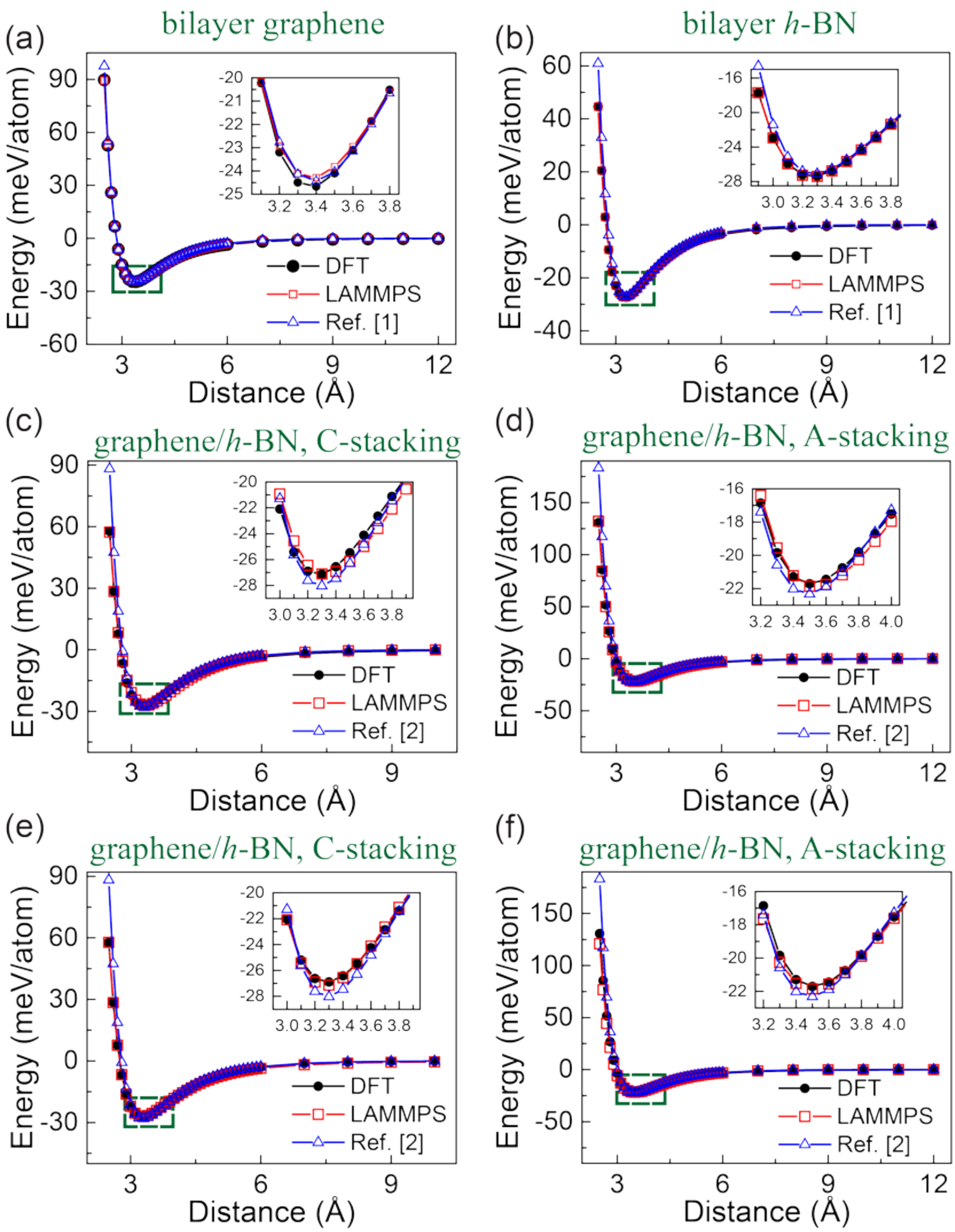

***Figure S1**: Binding energy curves of the laterally periodic bilayer structures of (a) graphene/graphene, (b) h-BN/h-BN, (c),(e) C-stack graphene/h-BN, (d),(f) A-stack graphene/h-BN. The results presented in panels (c) and (d) are calculated with the first set parameters (Table S1) and those presented in panels (e) and (f) are calculated with the second set parameters (Table S2). The reported energies are measured relative to the infinitely separated bilayer value and are normalized by the total number of atoms per unit-cell. The insets provide a zoom-in on the equilibrium interlayer separation region.*

The refined parameters proposed herein provide a satisfactory agreement with the reference binding energy curve within the long-range, near-equilibrium, and sub-equilibrium interlayer separation regimes. This improves upon our previous parameterizations, which shows large deviations in the sub-equilibrium region.[1, 2] However, we note that the reliability of the reference DFT calculations in the sub-equilibrium region, which is relevant for high pressure and tribological calculations, remains unclear. Hence, our fitting procedure mainly demonstrates the ability to obtain good agreement with reference data across the entire interlayer separation range. Nevertheless, in order to obtain reliable sub-equilibrium ILP results accurate reference data for this region should be provided.

An improved agreement with the reference data is also found for the finite homogenous (Figure S2) and heterogeneous (Figure S3) dimers, with the exception of Borazine. This is due to the weighting technique adopted during the fitting procedure, which gives less importance to the smaller systems (see Section 1).

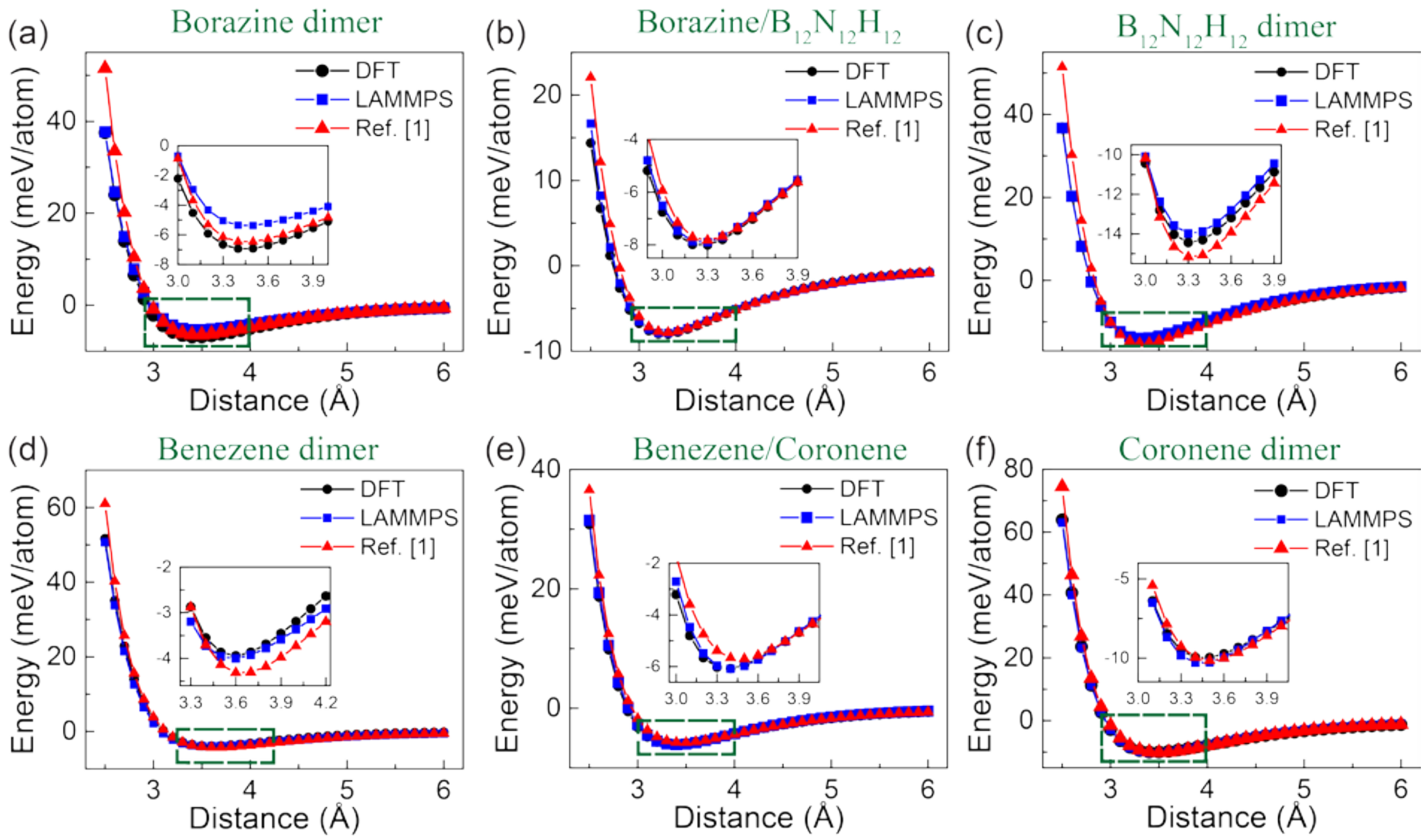

***Figure S2**: Binding energy curves calculated for the finite homogenous dimers of (a) Borazine, (b) Borazine/$B_{12}N_{12}H_{12}$, (c) $B_{12}N_{12}H_{12}$, (d) Benzene, (e) Benzene/Coronene, and (f) Coronene. The reported energies are measured relative to the infinitely separated dimer value and are normalized by the total number of atoms per unit-cell. The insets provide a zoom-in on the equilibrium interlayer separation region. Here, the parameters presented in Table S1 are used. Similar results are obtained when using the parameters of Table S2.*

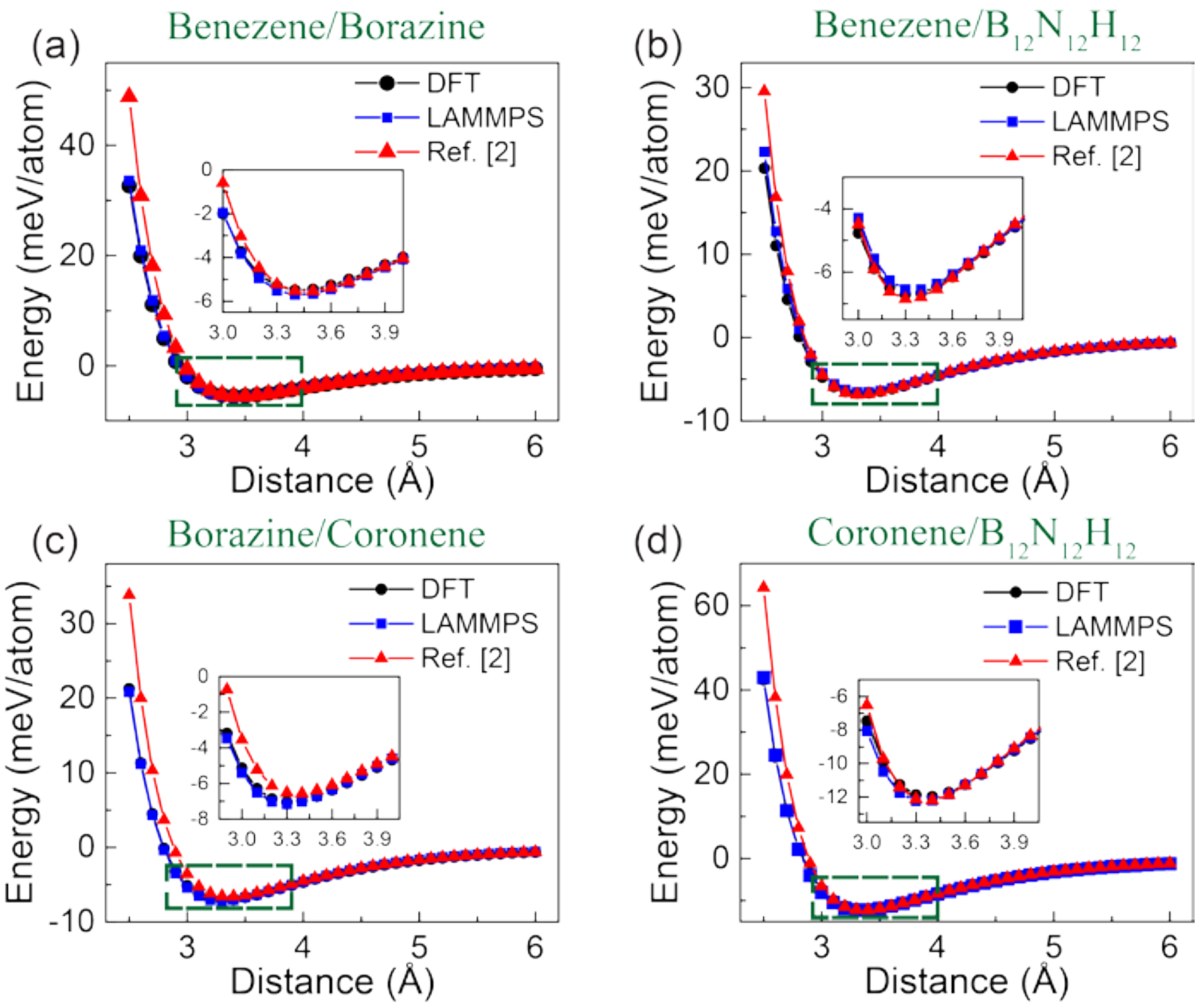

***Figure S3****: Binding energy curves calculated for the finite heterogeneous dimers of (a) Benzene/Borazine, (b) Benzene/ $B_{12}N_{12}H_{12}$, (c) Borazine/Coronene, and (d) Coronene/$B_{12}N_{12}H_{12}$. The reported energies are measured relative to the infinitely separated dimer value and are normalized by the total number of atoms per unit-cell. The insets provide a zoom-in on the equilibrium interlayer separation region. Here, the parameters presented in Table S1 are used. Similar results are obtained when using the parameters of Table S2.*

## 2.2 Sliding Energy Surfaces

A major advantage of the anisotropic ILP over isotropic pairwise potentials, such as Lennard-Jones and Morse potentials, is its ability to simultaneously capture both the interlayer binding and sliding energy surfaces of layered materials junctions.[1-4] This is demonstrated in Figure S4, where the ILP sliding energy surfaces obtained using the parameters of Table S1 for all the periodic structures are compared to the reference DFT data. The first and second rows in Figure S4 present the sliding energy surfaces of graphene/graphene, *h*-BN/*h*-BN and graphene/*h*-BN calculated using DFT and LAMMPS, respectively. The differences between the ILP and reference sliding data are presented in the third row of Figure S4. The largest deviation of ~1.5 meV/atom occurs for the heterogeneous graphene/*h*-BN junction. This deviation can be further reduced by using the parameters of Table S2 leading to a maximal deviation of ~0.6 meV/atom for the graphene/*h*-BN heterojunction as shown in Figure S5.

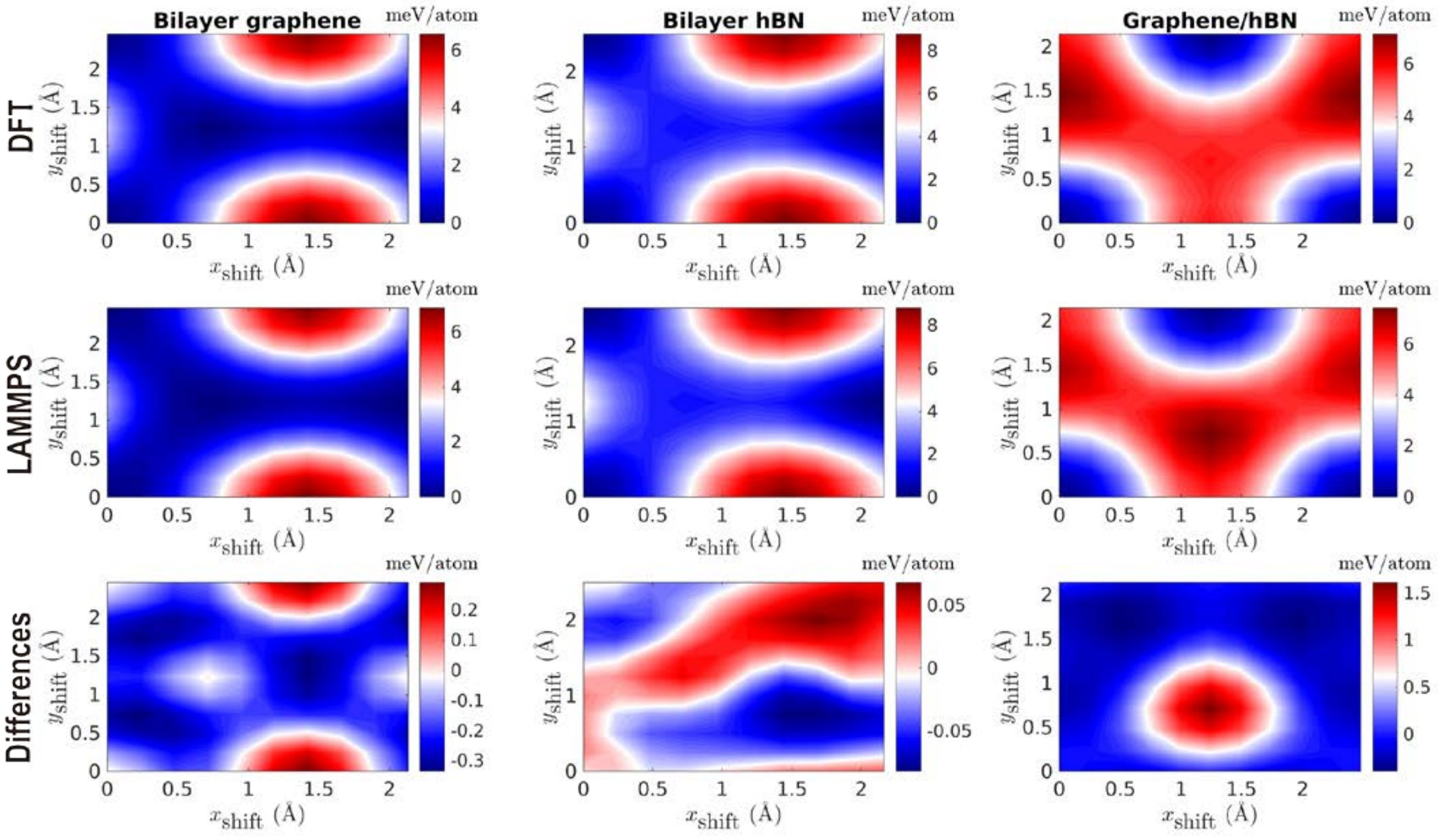

***Figure S4****: Sliding energy surfaces of the various periodic structures considered. The first and second rows present the sliding energy surface of graphene/graphene, h-BN/h-BN and graphene/h-BN bilayers calculated using dispersion augmented DFT and the LAMMPS implementation of the refined ILP, respectively. The third row presents their differences. The parameters of Table S1 are used in the ILP calculations.*

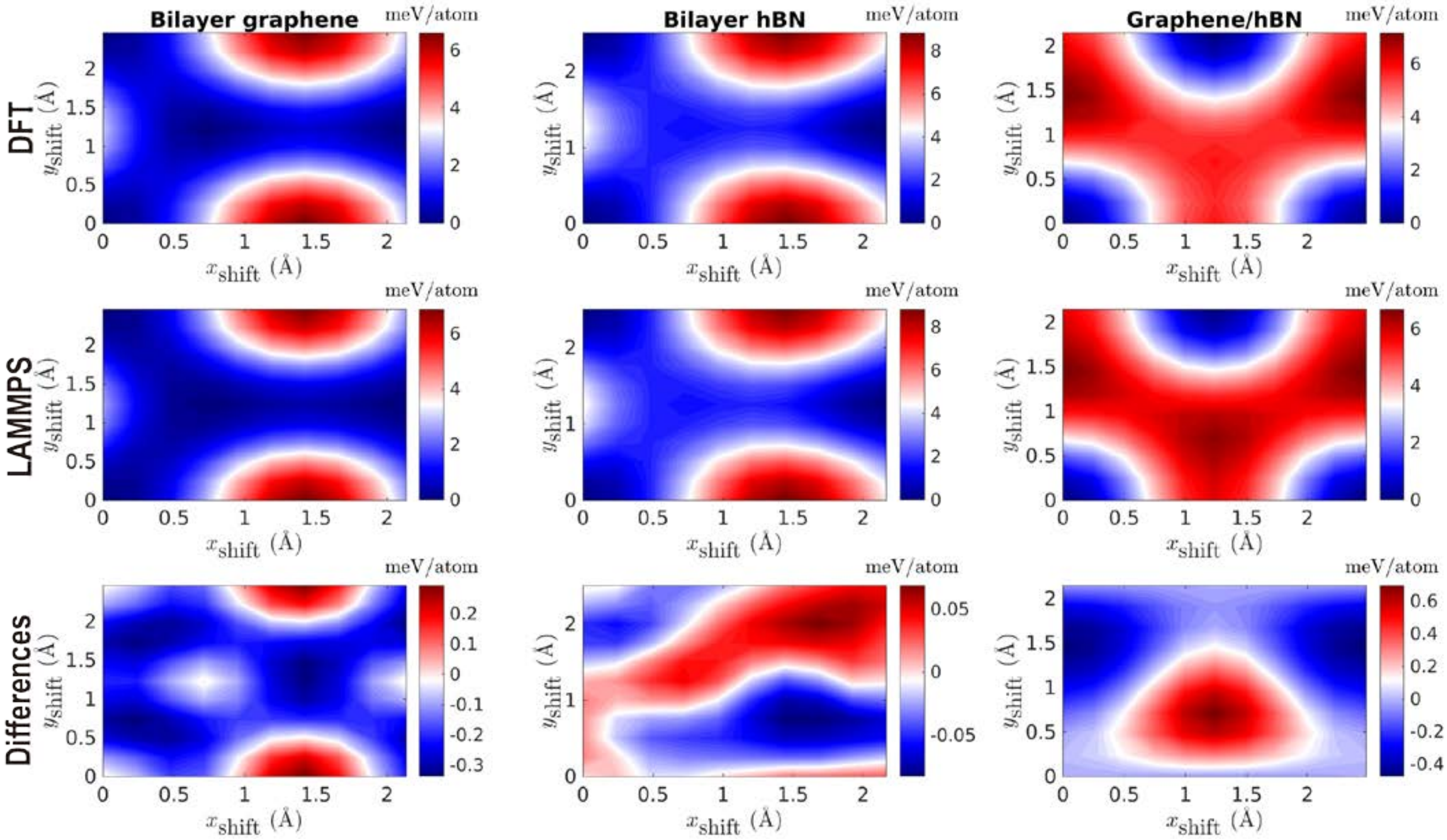

***Figure S5***: *Sliding energy surfaces of the various periodic structures considered. The first and second rows present the sliding energy surface of graphene/graphene, h-BN/h-BN and graphene/h-BN bilayers calculated using dispersion augmented DFT and the LAMMPS implementation of the refined ILP, respectively. The third row presents their differences. The parameters of Table S2 are used in the ILP calculations.*

## 2.3 Binding Energy Curves and Sliding Energy Surfaces Obtained Using the KC Potential

Figure S6 illustrates the refined KC potential benchmark tests for homogenous graphene bilayer. The refined parameters proposed herein provide a satisfactory agreement with the reference binding energy curve within the long-range, near-equilibrium, and sub-equilibrium interlayer separation regimes. This improves upon the original parameterizations for KC potential,[4] which shows larger deviations near equilibrium. Figure S6 e-f presents the differences of the sliding energy surfaces of bilayer graphene, between the original and refined KC potential parameterizations and the DFT reference data, respectively. The corresponding largest absolute deviations are ~0.6 and ~0.06 meV/atom.

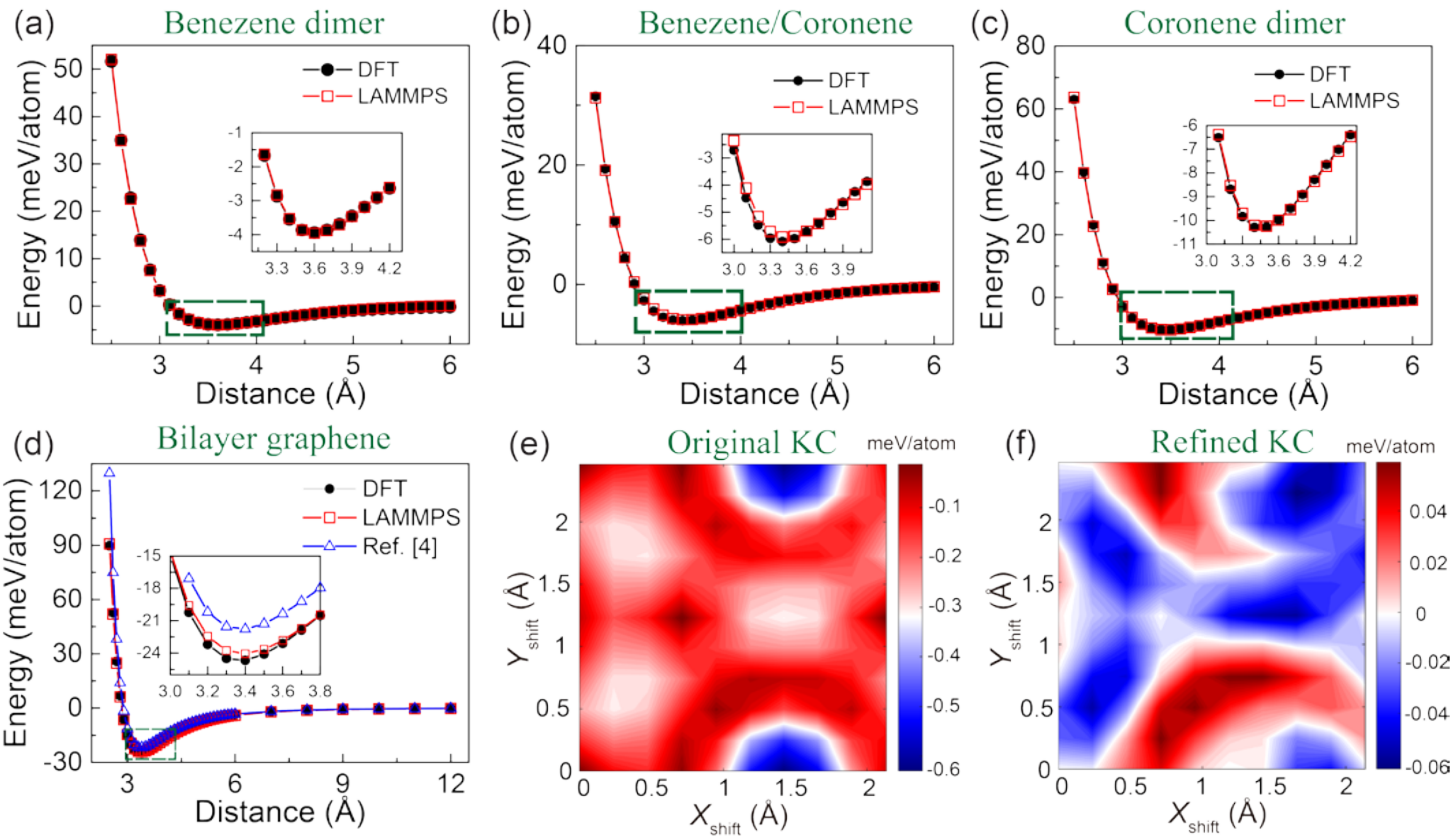

***Figure S6**: Benchmark tests for the KC potential. Binding energy curves calculated for the finite homogenous dimers of (a) Benzene, (b) Benzene/Coronene, (c) Coronene and for (d) periodic bilayer graphene. Energies are reported relative to the infinitely separated dimer value and are normalized by the total number of atoms per unit-cell. The insets provide a zoom-in on the equilibrium interlayer separation region. (e) bilayer graphene sliding energy surface difference between the LAMMPS implementation of the original KC potential and dispersion augmented DFT, (f) same as (e) but for refined KC potential. The parameters appearing in Table S3 are used herein.*

### 3. ILP Parameters Sensitivity Test

In order to check the sensitivity of the friction force results reported in the main text to the choice of ILP parameter set, we compare in Figure S7 the length dependence of static and kinetic friction forces of the GNR/*h*-BN heterojunctions for the two sets of parameters presented in Table S1 (full red circles) and Table S2 (open blue squares). The two sets produce very similar results, indicating that under the simulations conditions used herein the friction forces are relatively insensitive to the corresponding differences between the interaction potentials.

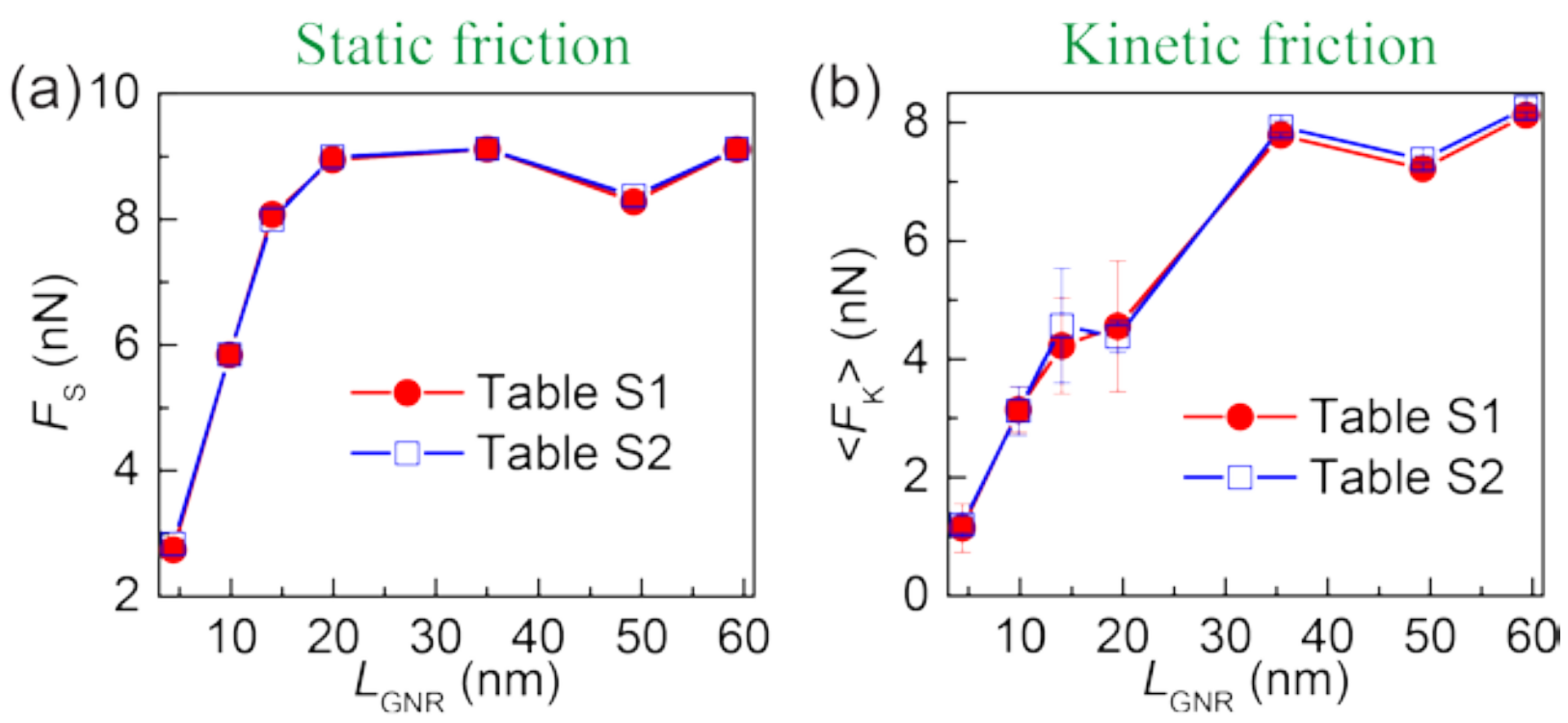

***Figure S7**: Sensitivity of the friction forces of a GNR/h-BN heterojunction towards the choice of ILP parameter set. Shown is the length dependence of the (a) static and (b) kinetic friction forces of the GNR calculated using the parameters presented in Table S1 (full red circles) and in Table S2 (open blue squares). The static friction force was evaluated from the maxima of the friction force traces. The kinetic friction force was calculated as* $\langle F_K \rangle = \langle K_{dr}(V_{dr}t - X_{edge}) \rangle$*, where* $\langle \cdot \rangle$ *denotes a steady-state time average. The statistical errors have been estimated using ten different trajectories, each averaged over a time interval of 1 ns.*

#### 4. Intra-layer Potential Sensitivity Test

In order to check the sensitivity of the friction force results reported in the main text to the choice of intra-layer potential we compare in Figure S8 the length dependence of static and kinetic friction forces of the GNR/*h*-BN heterojunctions obtained using the AIREBO[14] and the REBO[15] force-fields for graphene. Since the equilibrium intralayer C-C distances obtained with the AIREBO and REBO potential differ (1.3978 and 1.42 Å, respectively), we adjust the lattice constant of the rigid *h*-BN substrate accordingly to get the same lattice mismatch of 1.8 %. The two intra-layer terms produce very similar results indicating that under the simulation conditions used herein the friction forces are relatively insensitive to the choice of intra-layer potential. We note that in all simulations presented in the main text, the AIREBO potential has been used.

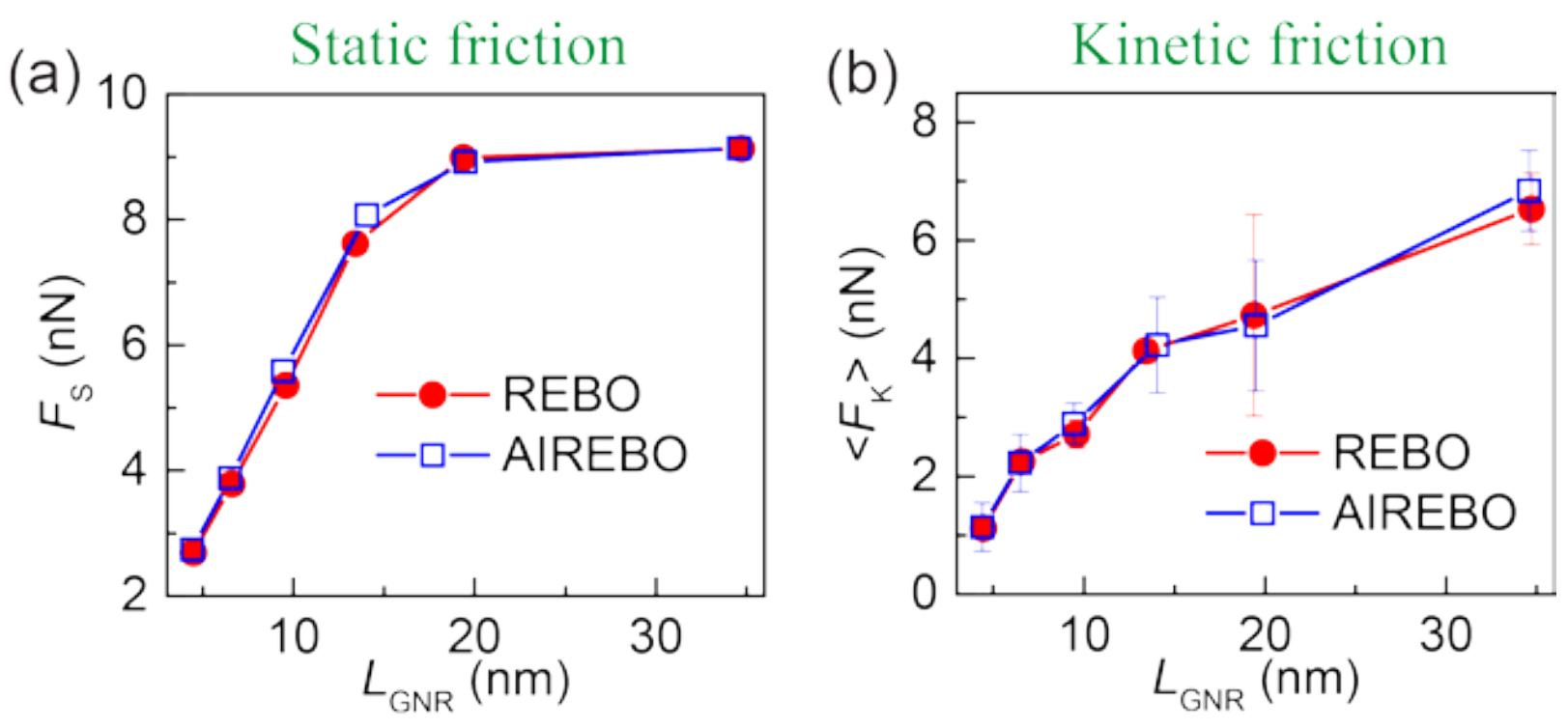

***Figure S8***: *Sensitivity of the friction forces of a GNR/h-BN heterojunction towards the choice of intra-layer potential. Shown is the length dependence of the (a) static and (b) kinetic friction forces calculated using the REBO (full red circles) and AIREBO (open blue squares) potentials. The statistical errors have been estimated as in Figure S7.*

**5. Damping Coefficient Sensitivity Test**

In order to check the sensitivity of the friction force results reported in the main text to the choice of damping coefficients (see eq 1 of the main text) we compare in Figure S9 the length dependence of static and kinetic friction forces of the GNR/graphene homogenous junctions obtained using three different values of $\eta^0$ spanning two orders of magnitude around the value adopted in the main text, $\eta^0 = 0.1, 1.0, 10.0\ \text{ps}^{-1}$. While in general we obtained similar qualitative trends, a somewhat increased friction is observed for the highest value considered. This is due to the increasing contribution of the viscous-like friction term of eq 1 of the main text. Noting that in typical experiments the pulling velocities are several orders of magnitude lower than those that can be practically simulated, the contribution of viscous-like friction in the simulation is irrelevant for the interpretation of experimental data.

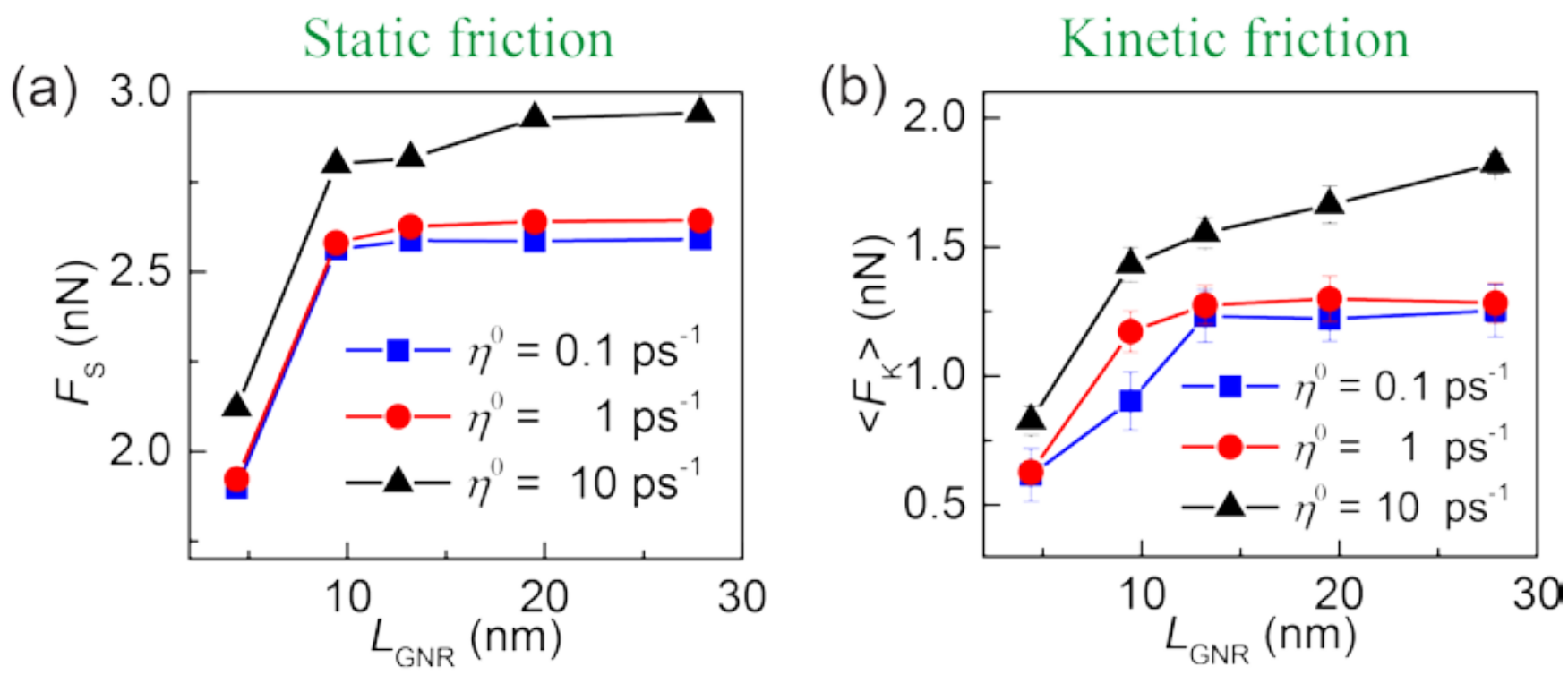

***Figure S9***: *Sensitivity of the friction forces of a GNR/graphene homogenous junction towards the choice of damping coefficients. Shown is the length dependence of the (a) static and (b) kinetic friction forces calculated using three values of the damping coefficients:* $\eta^0 = 0.1\ ps^{-1}$ *(full blue squares)* $\eta^0 = 1.0\ ps^{-1}$ *(full red circles), and* $\eta^0 = 10.0\ ps^{-1}$ *(full black triangles). The statistical errors have been estimated as in Figure S7.*

## 6. Propagation Time-Step Sensitivity Test

In order to check the sensitivity of the friction force results reported in the main text to the choice of propagation time-step we compare in Figure S10 the friction force traces of a 4.5 nm GNR sliding atop a graphene substrate obtained using a time-step of 1 fs (as in the main text) and 0.25 fs. We find that, despite the presence of light hydrogen atoms, a time-step of 1 fs is sufficient to provide converged results.[16]

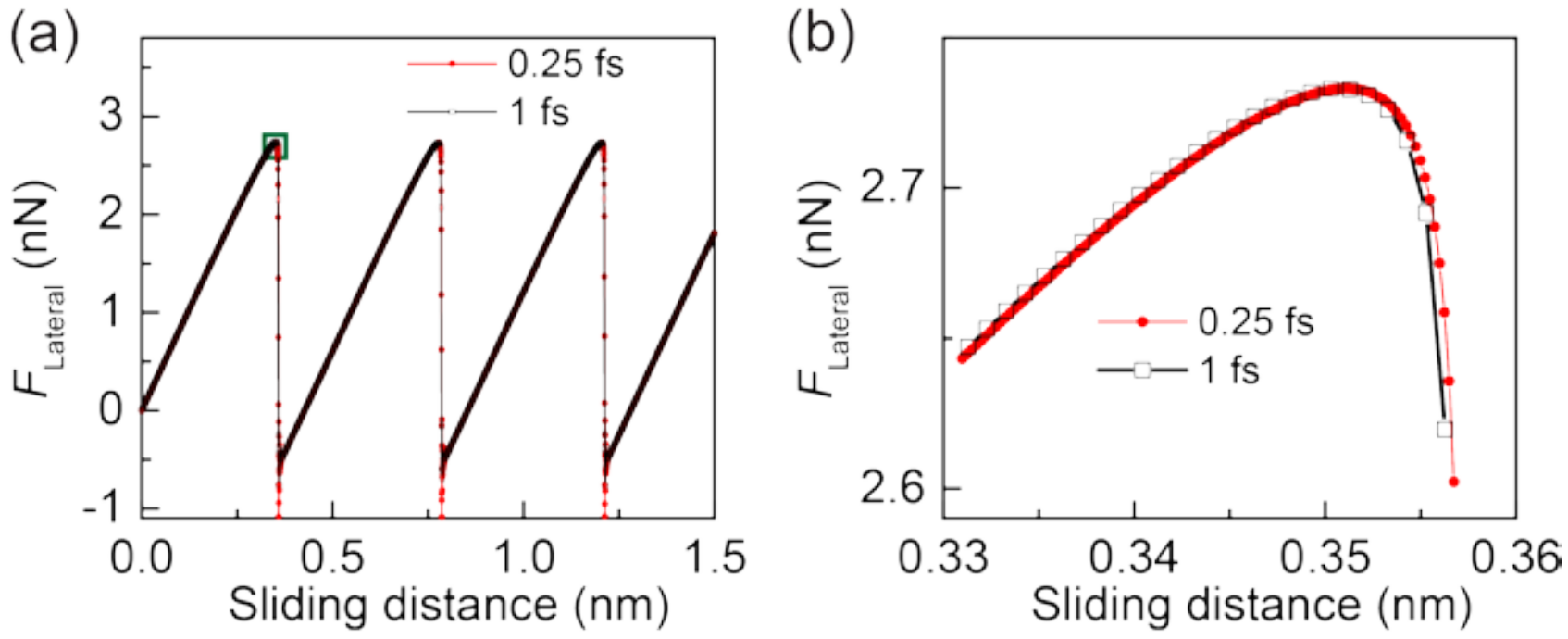

***Figure S10***: *Sensitivity of the friction force traces of a 4.5 nm GNR sliding atop a graphene substrate towards the choice of propagation time-step. Shown is (a) the full steady-state friction force trace and (b) a zoom-in on the peak region obtained using a time step of 1 fs (open black squares) and 0.25 fs (full red circles).*

## 7. Temperature Sensitivity Test

In order to check the sensitivity of the friction force results reported in the main text to the simulated temperature we compare in Figure S11 the length dependence of static and kinetic friction forces of the GNR/graphene homogeneous junctions obtained at zero temperature (as in the main text) and at room temperature. The latter simulations were performed using a Langevin thermostat applied to all slider atoms. The pulling velocity and stiffness of the springs are the same as that in the main text. Prior to the friction simulations, the system was equilibrated at 300 K for 400 ps with a time step of 0.5 fs. The results obtained at both temperatures show similar qualitative friction length dependence with somewhat reduced friction at room temperature, which results from thermally assisted barrier crossing.[17] This indicates that the mechanism leading to the unique friction force length dependence in this system remains valid also at room temperature.

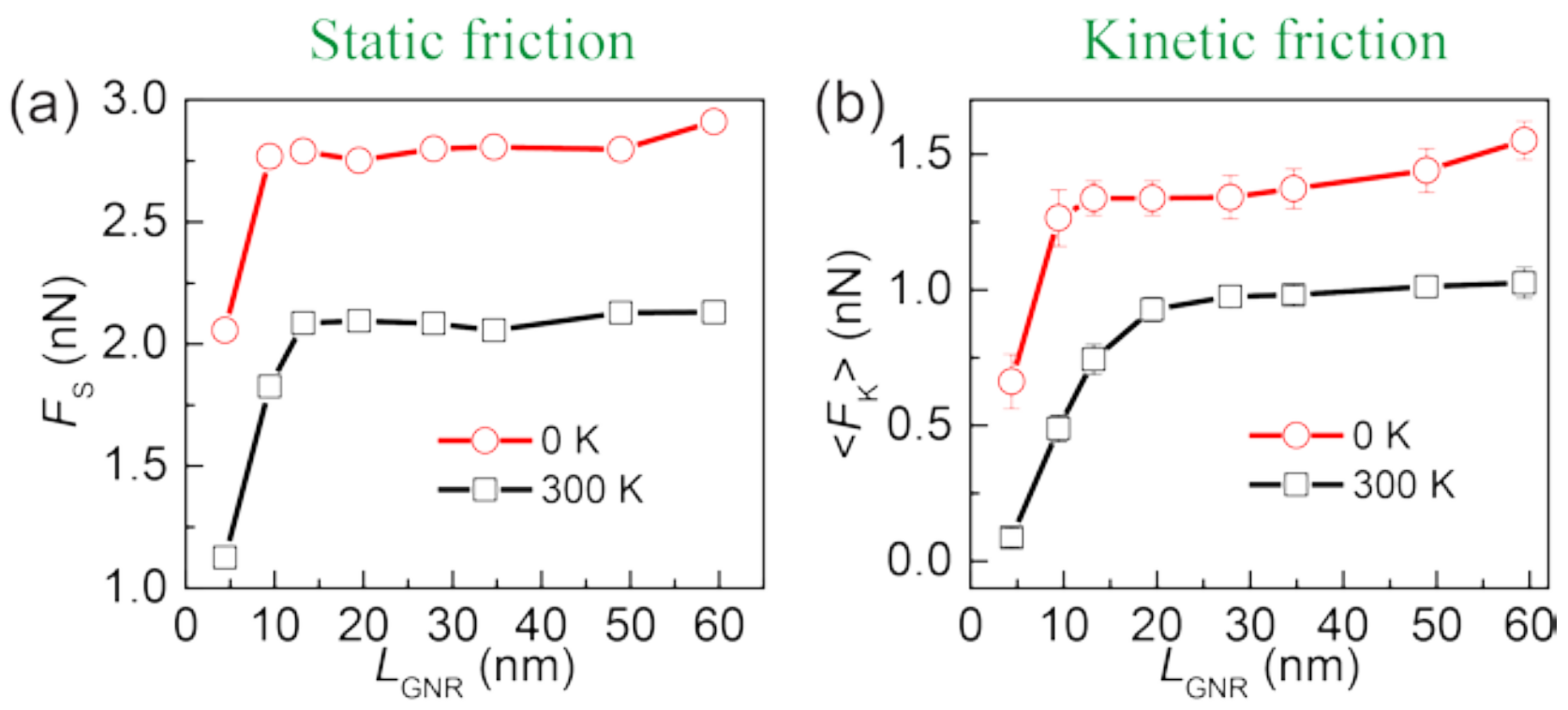

***Figure S11***: *Sensitivity of the friction forces of a GNR/graphene homogeneous junction towards the simulated temperature. Shown is the length dependence of the (a) static and (b) kinetic friction forces of the GNR length calculated at 0 K (open red circles) and 300 K (open black squares). Static and kinetic friction forces have been calculated as in Figure S7 above. The statistical errors have been estimated using ten different trajectories, each averaged over a time interval of 1.5 ns.*

## 8. Theoretical Estimation of the Characteristic Stress Decay Length

As mentioned in the main text, for commensurate contacting surfaces such as the aligned GNR/graphene interface the shear-induced stress distribution across the GNR can be described by a simplified one-dimensional model.[18] The characteristic stress decay length predicted by this model is given by $L_c = L_{GNR}\sqrt{K_{GNR}/K_{Interface}}$, where $L_{GNR}$ is the length of the GNR, $K_{Interface}$ is the interfacial shear stiffness between the GNR and the substrate and $K_{\mathrm{GNR}} = Ehb/L_{\mathrm{GNR}}$ is the in-plane stiffness of the GNR. Here, $E$, $h$, and $b$ are the Young's modulus, the thickness, and width of the GNR, respectively.

The interfacial shear stiffness, $K_{\mathrm{Interface}}$, has been evaluated by shifting the fully relaxed GNR rigidly over the graphene surface along the aligned sliding direction and fitting the deepest well obtained along the sliding potential energy curve to a parabola. As expected for commensurate contacts, this stiffness grows linearly with the GNR length (see Figure S12 a). To evaluate the GNR's in-plane stiffness we adopted the values $Eh = 26.6\ \mathrm{eV/Å^2}$ and $b$ = 0.726 nm.[19] A fit of the data reported in Figure S12a (see red line) yielded $K_{\mathrm{Interface}}/L_{\mathrm{GNR}} = 87.66 \pm 0.21\ \mathrm{eV/nm^3}$, giving $L_c \approx 4.6948 \pm 0.0056$ nm, somewhat larger than that obtained from fitting the MD simulation results (4.14 nm). The main reason for this discrepancy is that the theoretical estimation is based on a one-dimensional model, while the MD simulations allow for atomic motions in all directions. To prove this point, we performed additional simulation while freezing the atomic degrees of freedom perpendicular to the pulling direction. The resulting GNR stress-distribution color maps of the stress profile before the first slip event are illustrated in Figure S12 b and c. By fitting the stress profile with an exponential function, we obtained a characteristic stress decay length of 4.66 nm, in better agreement with the value predicted by the one-dimensional theory.

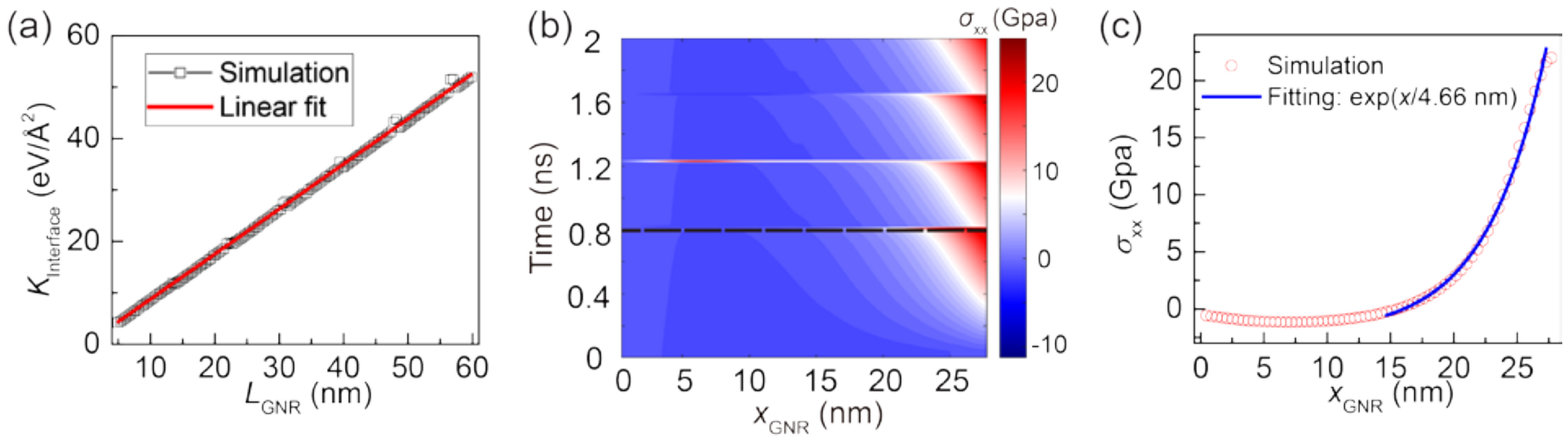

***Figure S12****: Estimation of the characteristic stress decay length. (a) The interfacial shear stiffness for the aligned homogeneous GNR/graphene junction as a function of the GNR length. The open black squares are simulation results and the red line is a linear fit. The sudden jumps in the simulation results are due to edge effects corresponding to a change in the local stacking of the leading edge atoms of the relaxed GNR relative to the graphene substrate. (b) Color maps showing the stress distribution along the GNR as a function of time for the aligned GNR/graphene junction. Here, all atoms within the GNR are constrained to move only along the pulling direction. (c) Open red circles show a cross section of the color map appearing in panel (b) at the onset of global sliding (dashed black line in panel (b)). The blue curve in panel (c) is an exponential fit with a characteristic stress decay length of 4.66 nm.*

## 9. Stacking Mode of the Leading GNR Edge Atoms for Heterogeneous GNR/*h*-BN Junctions

To explain the sharp jumps between two distinct values of static friction observed in GNR/*h*-BN heterojunctions with increasing ribbon length (Figure 2b in the main text), we show in Figure S13 the stacking mode of the leading edge of the ribbon for $L_{GNR}$=37.76 nm and $L_{GNR}$=48.08 nm, respectively, at the onset of a sliding event. We find that when the GNR exhibits even (odd) number of buckles, its leading edge atoms (marked in red) are positioned in an energetically (un)favorable stacking mode.

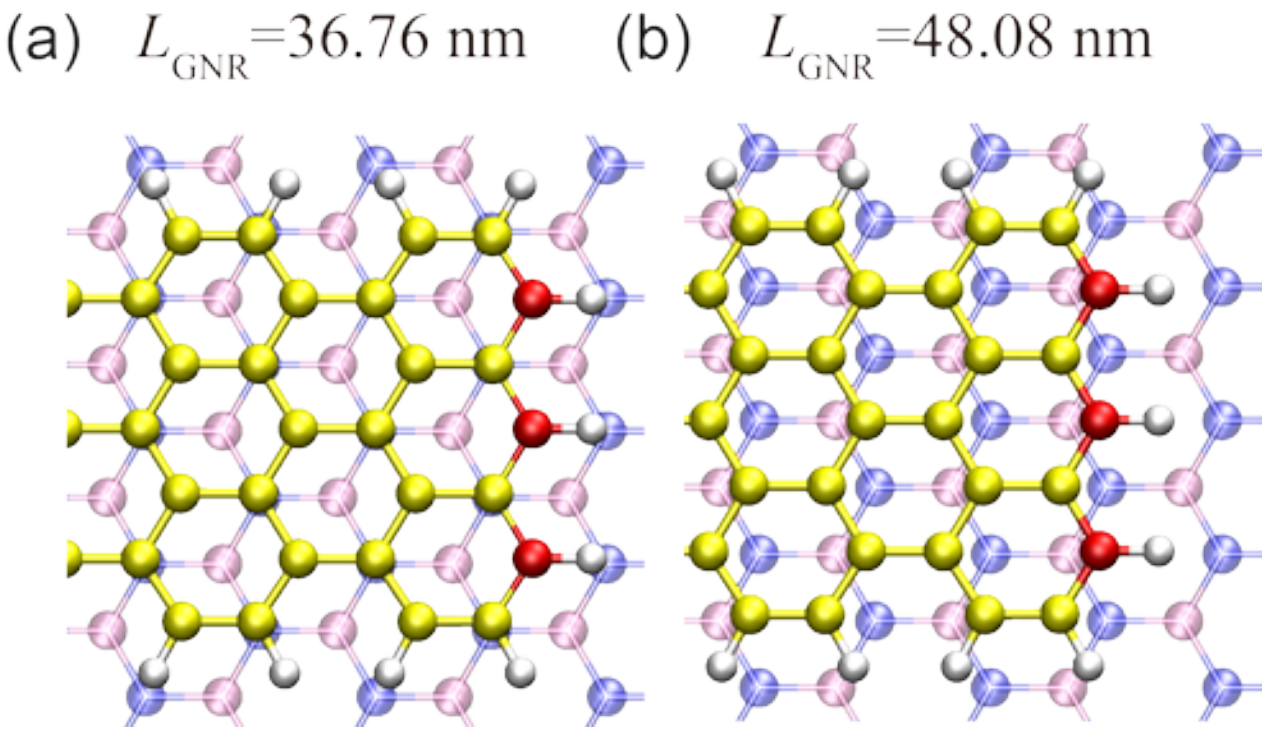

***Figure S13**: Stacking modes of the GNR leading edge atoms atop an h-BN substrate obtained for a ribbon length of (a) 37.76 nm and (b) 48.08 nm at the onset of a sliding event. Mauve, blue, yellow, and grey spheres represent boron, nitrogen, carbon, and hydrogen atoms, respectively. The leading edge atoms of the GNR are marked in red.*